\documentclass[a4paper,11pt]{article}
\pdfoutput=1 
\usepackage{jheppub} 
\usepackage[T1]{fontenc} 
\usepackage[utf8]{inputenc}
\usepackage{amsfonts}
\usepackage{graphics}
\usepackage{epsfig} 
\usepackage[english]{babel} 

\usepackage{bm}

\usepackage{multirow}
\usepackage{array,booktabs}
\usepackage{calc}
\usepackage[table]{xcolor}
\usepackage{color} 
\usepackage{makecell} 
\definecolor{aqua}{rgb}{0.0, 1.0, 1.0}
\definecolor{babyblue}{rgb}{0.54, 0.81, 0.94}
\definecolor{beaublue}{rgb}{0.74, 0.83, 0.9}
\definecolor{blizzardblue}{rgb}{0.67, 0.9, 0.93}
\definecolor{cyan}{rgb}{0.0, 1.0, 1.0}
\newcolumntype{?}{!{\vrule width 0.8pt}}

\newcommand{\RN}[1]{
  \textup{\uppercase\expandafter{\romannumeral#1}}%
}

\title{\boldmath Search for heavy neutral CP-even Higgs within Type-\RN{4} 2HDM at a future linear collider} 
\author[]{Majid Hashemi and Gholamhossein Haghighat}
\affiliation[]{Physics Department, College of Sciences, Shiraz University, \\ Shiraz, 71946-84795, Iran}
\emailAdd{hashemi$\_$mj$@$shirazu.ac.ir, hosseinhaqiqat$@$gmail.com}

\abstract{
In this paper, the production process $e^- e^+ \rightarrow A H$ is analyzed in the context of the type \RN{4} 2HDM and the question of observability of a neutral CP-even Higgs boson $H$ at a linear collider operating at $\sqrt{s}=1$ TeV is addressed. The CP-odd Higgs is assumed to experience a gauge-Higgs decay as $A\rightarrow ZH$ with hadronic decay of $Z$ boson as the signature of signal events. The production chain is thus $e^+e^- \rightarrow AH \rightarrow ZHH \rightarrow jj\ell\ell\ell\ell$ where $\ell$ is a $\tau$ or $\mu$. Four benchmark points with different mass hypotheses are assumed for the analysis. The Higgs mass $m_H$ is assumed to vary within the range 150-300 GeV in increments of 50 GeV. The anti-$k_t$ algorithm is used to perform the jet reconstruction. Results indicate that the neutral CP-even Higgs $H$ is observable through this production mechanism using the di-muon invariant mass distribution with possibility of mass measurement. The corresponding signal significances exceed $5\sigma$ at integrated luminosity of 3000 $fb^{-1}$. 
 }     

\begin{document}
\maketitle 
\flushbottom
 
\section{Introduction}
The standard model of elementary particles has been confirmed by a large number of experimental tests and its success has attracted attention on possible extensions of the theory which may pave the way for a solution to the present serious and challenging problems in physics. The discovery of the first elementary scalar particle, the Higgs boson~\cite{HiggsObservationCMS,HiggsObservationATLAS}, which was a prediction of the Higgs mechanism  \cite{SM1,SM2,SM3,SM4,SM5,SM6}, has strengthened the idea that multiple Higgs bosons may exist in nature and the discovered SM Higgs may be only one of them. In the SM, the simplest possible scalar structure is assumed. However, supersymmetry~\cite{MSSM1}, axion models~\cite{axionModels}, the SM inability to explain the baryon asymmetry of the universe \cite{SMinabilityToExplainBaryonAsymmetry}, etc., have been some important motivations behind the idea of extending the SM by adding another $SU(2)$ Higgs doublet. Assuming two $SU(2)$ Higgs doublets instead of a single doublet forms one of the simplest extensions of the SM, i.e., the two-Higgs-doublet model (2HDM)~\cite{2hdm_TheoryPheno,2hdm1,2hdm2,2hdm3,2hdm4_CompositeHiggs,2hdm_HiggsSector1,2hdm_HiggsSector2}. 

Depending on how different types of fermions couple to Higgs doublets, one can divide 2HDM into 5 types, four of which with natural flavour conservaion and the fifth with flavour-changing neutral currents (FCNCs). Since no evidence for FCNC has been observed in nature, types with flavour conservation have been more interesting. In comparison with the SM, probing the entire parameter space of 2HDMs takes much longer because of the large number of free parameters in these models. 2HDMs predict more than one Higgs particle, one of which is expected to be the same as the SM Higgs. Other than the SM-like Higgs boson $h$, two other neutral Higgs bosons $A$ and $H$, and two charged Higgs bosons $H^\pm$ are predicted in 2HDMs. In this work the observability of the scalar CP-even neutral Higgs $H$ is studied.
 
The production process $e^+ e^- \rightarrow AH\rightarrow ZHH$ followed by the decays $Z\rightarrow jj$ and $HH\rightarrow \tau \tau \mu \mu \ or\ \mu \mu \tau \tau$ is assumed in this work to take advantage of the clear signal that leptonic decays can provide at linear colliders. From the four types of the 2HDM which conserve flavour, the type \RN{4} is chosen as the theoretical framework to enhance the Higgs leptonic decay at high $\tan\beta$ by utilizing the key role $\tan\beta$ plays~\cite{tanbsignificance}. This enhancement is caused by the appearance of $\cot\beta$ and $\tan\beta$ factors in Higgs-quark and Higgs-lepton couplings respectively in the Yukawa lagrangian of the type \RN{4} 2HDM. $\tau$-jets in the final decay products are identified by performing a $\tau$-tagger algorithm and it is expected that the di-muon invariant mass distribution will provide an observable signal on top of the background. 

Contrary to the Minimal Supersymmetric Standard Model (MSSM)~\cite{MSSM1,MSSM2,MSSM3,MSSM4} which constrains the Higgs masses, Higgs masses of the 2HDM are allowed to have arbitrary values in general. Therefore, 2HDMs provide a wider mass parameter space. In this work, four benchmark points in the Higgs mass parameter space are defined and a simulation is performed for each one to assess the observability of the neutral Higgs $H$. It will be shown that the signal is observable and the $H$ mass reconstruction is possible for all of the four assumed benchmark points. In~\cite{Kanemura} a similar signal in the context of type \RN{4} 2HDM with multi-tau-lepton signature has also been studied at LHC with promising results. 

\section{Two-Higgs-doublet model}
Contrary to the SM, where the diagonalized mass matrix leads to the diagonalized Yukawa interactions, the mass matrix is not diagonalizable in a general 2HDM and thus 2HDM contains FCNCs in general. Avoiding the severe difficulties arising from FCNCs, Paschos-Glashow-Weinberg theorem~\cite{2hdm2,naturalflavorcons.} states that FCNCs will be removed if all fermions with the same quantum numbers couple exactly to one of the two Higgs doublets. Following this theorem and assuming that the Higgs-fermion couplings follow from table \ref{coupling}, the four types of 2HDM with natural flavour conservation are produced. 
\begin{table}[h]
\normalsize
\fontsize{11}{7.2} 
    \begin{center}
         \begin{tabular}{ >{\centering\arraybackslash}m{.3in} ? >{\centering\arraybackslash}m{1.4in} ? >{\centering\arraybackslash}m{1.4in} ?>{\centering\arraybackslash}m{1.4in} ? >{\centering\arraybackslash}m{1.4in} ?}
& \cellcolor{blizzardblue}{Up-type quarks couple to} & \cellcolor{blizzardblue}{Down-type quarks couple to} & \cellcolor{blizzardblue}{Leptons couple to} \parbox{0pt}{\rule{0pt}{1ex+\baselineskip}}\\ \Xhline{6\arrayrulewidth}
 \cellcolor{blizzardblue}{$\RN{1}$} &$\Phi_2$ &$\Phi_2$ &$\Phi_2$ \parbox{0pt}{\rule{0pt}{1ex+\baselineskip}}\\ \Xhline{2\arrayrulewidth}
 \cellcolor{blizzardblue}{$\RN{2}$} &$\Phi_2$ &$\Phi_1$ &$\Phi_1$ \parbox{0pt}{\rule{0pt}{1ex+\baselineskip}}\\ \Xhline{2\arrayrulewidth}
\cellcolor{blizzardblue}{$\RN{3}$} &$\Phi_2$ &$\Phi_2$ &$\Phi_1$  \parbox{0pt}{\rule{0pt}{1ex+\baselineskip}}\\ \Xhline{2\arrayrulewidth}
\cellcolor{blizzardblue}{$\RN{4}$} &$\Phi_2$ &$\Phi_1$ &$\Phi_2$  \parbox{0pt}{\rule{0pt}{1ex+\baselineskip}}\\ \Xhline{2\arrayrulewidth}
  \end{tabular}
\caption{The Higgs doublet to which different fermions couple in the four types of the 2HDM. \label{coupling}}
  \end{center}
\end{table}

Following the coupling strategy of table \ref{coupling}, the Higgs sector of the 2HDM as introduced in~\cite{2hdm_TheoryPheno} encodes neutral and charged Higgs interactions through the Yukawa Lagrangian 
\begin{equation}
\begin{aligned}
\mathcal{L}_{\ Yukawa}\ =\ & -\ \sum_{f=u,d,\ell}\ \dfrac{m_f}{v}\ \Big(\xi_{h}^{f}\bar{f}fh\ +\ \xi_{H}^{f}\bar{f}fH\ -\ i\xi_{A}^{f}\bar{f}\gamma_5fA \Big)\\
&-\ \Bigg\{\dfrac{\sqrt{2}V_{ud}}{v}\bar{u}\ \big(m_u\xi_A^uP_L\ +\ m_d\xi_A^dP_R\big)\ dH^+\ +\ \dfrac{\sqrt{2}m_\ell\xi_A^\ell}{v}\overline{\nu_L}\ell_RH^+\ +\ H.c. \Bigg\}
\label{yukawa1}
\end{aligned}
\end{equation}
where $h,H,A$ and $H^+$ are SM-like Higgs, scalar CP-even neutral Higgs, pseudoscalar neutral Higgs and charged Higgs fields, $u,d$ and $\ell$ are up-type quark, down-type quark and lepton fields, $P_{L/R}$ are projection operators for left-/right-handed fermions, and finally the factors $\xi$, as presented in table \ref{xi} for different types, are factors expressed in terms of trigonometric functions of the parameters $\alpha$ and $\beta$.
\begin{table}[h]
\normalsize
\fontsize{11}{7.2} 
    \begin{center}
         \begin{tabular}{ >{\centering\arraybackslash}m{.3in} ? >{\centering\arraybackslash}m{1in} ? >{\centering\arraybackslash}m{1in} ?>{\centering\arraybackslash}m{1in} ? >{\centering\arraybackslash}m{1in} ? >{\centering\arraybackslash}m{1in} ?}
& \cellcolor{blizzardblue}{$\RN{1}$} & \cellcolor{blizzardblue}{$\RN{2}$} & \cellcolor{blizzardblue}{$\RN{3}$} & \cellcolor{blizzardblue}{$\RN{4}$ } \parbox{0pt}{\rule{0pt}{1ex+\baselineskip}}\\ \Xhline{6\arrayrulewidth}
 \cellcolor{blizzardblue}{$\xi_h^u$} &$\cos\alpha/\sin\beta$ &$\cos\alpha/\sin\beta$ &$\cos\alpha/\sin\beta$  &$\cos\alpha/\sin\beta$  \parbox{0pt}{\rule{0pt}{1ex+\baselineskip}}\\ \Xhline{2\arrayrulewidth}
 \cellcolor{blizzardblue}{$\xi_h^d$} &$\cos\alpha/\sin\beta$ &$-\sin\alpha/\cos\beta$ &$-\sin\alpha/\cos\beta$  &$\cos\alpha/\sin\beta$  \parbox{0pt}{\rule{0pt}{1ex+\baselineskip}}\\ \Xhline{2\arrayrulewidth}
 \cellcolor{blizzardblue}{$\xi_h^\ell$} &$\cos\alpha/\sin\beta$ &$-\sin\alpha/\cos\beta$ &$\cos\alpha/\sin\beta$   &$-\sin\alpha/\cos\beta$  \parbox{0pt}{\rule{0pt}{1ex+\baselineskip}}\\ \Xhline{2\arrayrulewidth}
 \cellcolor{blizzardblue}{$\xi_H^u$} &$\sin\alpha/\sin\beta$ &$\sin\alpha/\sin\beta$ &$\sin\alpha/\sin\beta$  &$\sin\alpha/\sin\beta$  \parbox{0pt}{\rule{0pt}{1ex+\baselineskip}}\\ \Xhline{2\arrayrulewidth}
 \cellcolor{blizzardblue}{$\xi_H^d$} &$\sin\alpha/\sin\beta$ &$\cos\alpha/\cos\beta$ &$\cos\alpha/\cos\beta$  &$\sin\alpha/\sin\beta$  \parbox{0pt}{\rule{0pt}{1ex+\baselineskip}}\\ \Xhline{2\arrayrulewidth}
 \cellcolor{blizzardblue}{$\xi_H^\ell$} &$\sin\alpha/\sin\beta$ &$\cos\alpha/\cos\beta$ &$\sin\alpha/\sin\beta$  &$\cos\alpha/\cos\beta$  \parbox{0pt}{\rule{0pt}{1ex+\baselineskip}}\\ \Xhline{2\arrayrulewidth}
 \cellcolor{blizzardblue}{$\xi_A^u$} &$\cot\beta$ &$\cot\beta$ &$\cot\beta$  &$\cot\beta$  \parbox{0pt}{\rule{0pt}{1ex+\baselineskip}}\\ \Xhline{2\arrayrulewidth}
 \cellcolor{blizzardblue}{$\xi_A^d$} &$-\cot\beta$ &$\tan\beta$ &$\tan\beta$  &$-\cot\beta$  \parbox{0pt}{\rule{0pt}{1ex+\baselineskip}}\\ \Xhline{2\arrayrulewidth}
 \cellcolor{blizzardblue}{$\xi_A^\ell$} &$-\cot\beta$ &$\tan\beta$ &$-\cot\beta$  &$\tan\beta$  \parbox{0pt}{\rule{0pt}{1ex+\baselineskip}}\\ \Xhline{2\arrayrulewidth}
  \end{tabular}
\caption{Factors $\xi_Y^X$ in different types of 2HDM. \label{xi}}
  \end{center}
\end{table}
As seen, different types use different couplings and therefore, exhibit different behaviours and collider phenomenology~\cite{2hdm_HiggsSector2}. The types \RN{3} and \RN{4} are also called ``flipped'' (or ``type Y'') and ``lepton-specific'' (or ``type X'') respectively. 

Taking the scalar neutral Higgs field $h$ as the SM-like Higgs field, the Yukawa interactions of the Higgs boson $h$ in 2HDM reduces to those of the SM by assuming $\sin(\beta-\alpha)=1$~\cite{2hdm_TheoryPheno}. Under this assumption, the neutral Higgs part of the Yukawa lagrangian takes the form~\cite{Barger_2hdmTypes} 
\begin{equation}
\begin{aligned}
v\ \mathcal{L}_{\ Yukawa}\ =\ & -\ \Big(\ m_d\ \bar{d}d\ +\ m_u\ \bar{u}u\ +\ m_\ell\ \bar{\ell}\ell\ \Big)\ h \\
      & +\ \Big(\ \rho^dm_d\ \bar{d}d\ +\ \rho^um_u\ \bar{u}u\ +\ \rho^\ell m_\ell\ \bar{\ell}\ell\ \Big)\ H \\
& +\ i\Big(-\rho^dm_d\ \bar{d}\gamma_5d\ +\ \rho^um_u\ \bar{u}\gamma_5u\ -\ \rho^\ell m_\ell\ \bar{\ell}\gamma_5\ell\ \Big)\ A
\label{yukawa2}
\end{aligned}
\end{equation}
where the $\rho$ factors of different types of 2HDM are presented in table \ref{rho}.
\begin{table}[h]
\normalsize
\fontsize{11}{7.2} 
    \begin{center}
         \begin{tabular}{ >{\centering\arraybackslash}m{.3in} ? >{\centering\arraybackslash}m{.6in} ? >{\centering\arraybackslash}m{.6in} ?>{\centering\arraybackslash}m{.6in} ? >{\centering\arraybackslash}m{.6in} ? >{\centering\arraybackslash}m{.6in} ?}
& \cellcolor{blizzardblue}{$\RN{1}$} & \cellcolor{blizzardblue}{$\RN{2}$} & \cellcolor{blizzardblue}{$\RN{3}$} & \cellcolor{blizzardblue}{$\RN{4}$} \parbox{0pt}{\rule{0pt}{1ex+\baselineskip}}\\ \Xhline{6\arrayrulewidth}
 \cellcolor{blizzardblue}{$\rho^d$} &$\cot{\beta}$ &$- \tan\beta$ &$-\tan\beta$ &$\cot\beta$ \parbox{0pt}{\rule{0pt}{1ex+\baselineskip}}\\ \Xhline{2\arrayrulewidth}
\cellcolor{blizzardblue}{$\rho^u$} &$\cot{\beta}$ &$\cot\beta$ &$\cot\beta$ &$\cot\beta$ \parbox{0pt}{\rule{0pt}{1ex+\baselineskip}}\\ \Xhline{2\arrayrulewidth}
\cellcolor{blizzardblue}{$\rho^\ell$} &$\cot{\beta}$ &$- \tan\beta$ &$\cot\beta$ &$-\tan\beta$ \parbox{0pt}{\rule{0pt}{1ex+\baselineskip}}\\ \Xhline{2\arrayrulewidth} 
 \end{tabular}
\caption{The $\rho$ factors of the Yukawa lagrangian in different types of the 2HDM. \label{rho}}
  \end{center}  
\end{table}

According to table \ref{rho}, the type \RN{1} is well-suited to low $\tan\beta$ studies because of the $\cot\beta$ factor appearing in its couplings. A study in the context of this type shows that the production of the pseudoscalar Higgs $A$ decaying into $ZH$ can be observed at LHC for certain $H$ decay modes~\cite{2HDMTypeI_LHC}.

\begin{figure}[h]
  \centering
  \includegraphics[width=.6\textwidth]{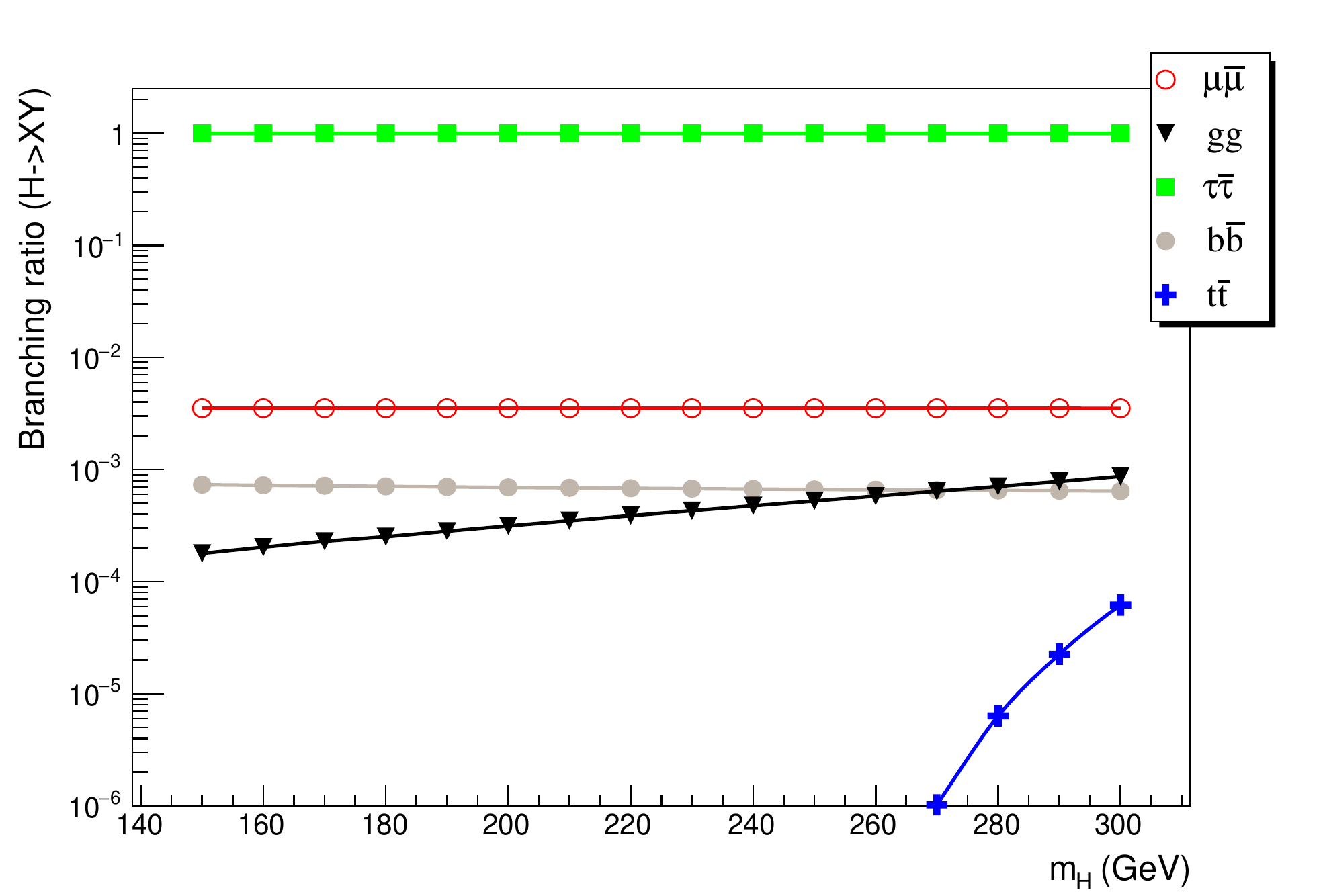} 
  \caption{Branching ratio of the Higgs boson $H$ different decay channels as a function of Higgs mass $m_H$. \label{HdecayBR}}
\end{figure}

The type \RN{4} (first discussed in~\cite{2hdmtype4-1,2hdmtype4-2}) also provides an interesting environment for studying leptonic decays of the neutral Higgs bosons $H$ and $A$, since the corresponding couplings are enhanced as $\tan\beta$ and the Higgs-quark couplings are suppressed at high $\tan\beta$, as seen in table \ref{rho}. That is the reason for searching for the $H$ leptonic decay through the di-muon invariant mass distribution in this work. Figure \ref{HdecayBR} shows the branching ratio of different $H$ decay channels at type \RN{4} 2HDM. As seen, the $\tau$ pair production channel has a larger branching ratio than di-muon channel, because of the larger $\tau$ lepton mass.

\section{Signal and background processes}
The type \RN{4} 2HDM is chosen as the theoretical framework in this work. Based on the features of this model, the process $e^+ e^- \rightarrow AH \rightarrow ZHH \rightarrow jj\tau\tau\mu\mu$ or $jj\mu\mu\tau\tau$ is defined as the signal process and a linear collider operating at an energy of 1 TeV is assumed.  The leptonic decay is chosen as the decay mode of the neutral Higgs $H$ so that the analysis benefits from the enhanced $H$ decay due to the $\tan\beta$ factor in the Higgs-lepton coupling. In addition, the leptonic decay mode is beneficial from the reconstruction efficiency aspect too. Since the lepton reconstruction efficiency at linear colliders is relatively high it is expected that the di-lepton (di-muon in this case) invariant mass will provide a clear signal. Despite using the muon pair to reconstruct the $H$ mass, one of the $H$ particles is defined to decay via decay mode $H\rightarrow \tau\tau$. The reason is that the branching ratio of the decay to a tau pair is so high ($BR_{H\rightarrow \tau\tau}=0.99$) suppressing the decay to muon pair to the level of few permil and the signal cross section would be very small if we choose muonic decay mode for both $H$ bosons.

As~\cite{L2HDMLHC} shows, a leptophilic neutral Higgs boson lighter than 140 GeV can be observed at 30 $fb^{-1}$ at LHC. In the present work, we focus on moderate and high masses region and assume four benchmark points (BPs) with different mass hypotheses. Table \ref{BPs} presents different parameters of the selected benchmark points. 
\begin {table}[h]
\begin{center}
\begin{tabular}{c?c?c?c?c?}
& \multicolumn{4}{ c? }{\cellcolor{blizzardblue}{Benchmark point}} \\ 
& \multicolumn{1}{ c? }{\cellcolor{blizzardblue}BP1} & \cellcolor{blizzardblue}BP2 & \cellcolor{blizzardblue}BP3 & \cellcolor{blizzardblue}BP4 \\ \Xhline{2\arrayrulewidth}
\multicolumn{1}{ c?  }{\cellcolor{blizzardblue}$m_{h}$} & \multicolumn{4}{ c? }{125} \\ \Xhline{2\arrayrulewidth}
\multicolumn{1}{ c?  }{\cellcolor{blizzardblue}$m_{H}$} & 150 & 200 & 250 & 300 \\ \Xhline{2\arrayrulewidth}
\multicolumn{1}{ c?  }{\cellcolor{blizzardblue}$m_{A}$} & 300 & 350 & 400 & 450 \\ \Xhline{2\arrayrulewidth}
\multicolumn{1}{ c?  }{\cellcolor{blizzardblue}$m_{H^\pm}$} & 300 & 350 & 400 & 450 \\ \Xhline{2\arrayrulewidth}
\multicolumn{1}{ c?  }{\cellcolor{blizzardblue}$\tan\beta$} & \multicolumn{4}{ c? }{10} \\ \Xhline{2\arrayrulewidth}
\multicolumn{1}{ c?  }{\cellcolor{blizzardblue}$\sin(\beta-\alpha)$} & \multicolumn{4}{ c? }{1} \\ \Xhline{2\arrayrulewidth}
\end{tabular}
\caption {Selected benchmark points. \label{BPs}}
\end{center}
\end {table}
As seen in table \ref{BPs}, the SM-like Higgs mass $m_h$, $\tan\beta$ and $\sin(\beta-\alpha)$ are assumed to be 125, 10 and 1 respectively for all of the selected points and the Higgs mass $m_H$ varies from 150 GeV for BP1 to 300 GeV for BP4 in increments of 50 GeV. 

The benchmark points are all checked to satisfy the constraints on $\rho=m_W^2(m_Z\cos\theta_W)^{-2}$ parameter which may deviate from its SM value in extended models like 2HDMs. The measurement performed at LEP \cite{2HDMrhoConstrainingMeasurement}, which is in excellent agreement with SM predictions, constrains the $\rho$ parameter in 2HDM~\cite{2HDMrhoConstraint-1,2HDMrhoConstraint-2}. It has been observed that degenerate Higgs boson masses produce negligible deviations of the $\rho$ parameter from the corresponding SM value \cite{2HDMdeltaRho}. Defining $\Delta\rho$ as the non-SM part of the $\rho$, the pseudoscalar and charged Higgs masses in the benchmark points are chosen to be the same so that $\Delta\rho$ is reduced to allowed values consistent with the provided constraints. 

The selected benchmark points are also checked to make sure that the resulting vacuum configurations are stable. The stability of a vacuum configuration is ensured by the positivity of the Higgs potential for asymptotically large values of the fields~\cite{2HDMHiggsPotentialPositivity}. Moreover, the selected points satisfy the constraints imposed by requiring perturbativity as well as tree-level unitarity for the scattering of Higgs bosons and electroweak gauge bosons~\cite{2HDMtreeLevelUnitarity-1,2HDMtreeLevelUnitarity-2,2HDMtreeLevelUnitarity-3,2HDMtreeLevelUnitarity-4}. All of these requirements are checked using 2HDMC 1.6.3~\cite{2hdmc1,2hdmc2} and are satisfied.

The current experimental limits on the pseudoscalar and charged Higgs masses are $m_{H^\pm}\geq78.6$ GeV and $m_A\geq93.4$ GeV, as shown in~\cite{lep1,lep2,lepexclusion1,lepexclusion2}. These constraints are obtained based on MSSM and cannot be applied to type \RN{4} 2HDM and thus in general, we are not required to respect these limits. However, since moderate and high masses region is our target in this work, the selected points are already consistent with these experimental limits.

Searches for heavy neutral Higgs boson at LHC have recently excluded the mass range $m_{A/H}=200-400$ GeV for $\tan\beta\geq5$~\cite{CMSNeutralHiggs,ATLASneutralHiggs}. However, this exclusion is also based on the MSSM and imposes no restriction on the mass spectrum space of this work, since MSSM Higgs-fermion couplings are different from those of the type \RN{4} 2HDM. Moreover, Higgs mass parameters in MSSM are not all free parameters like the mass parameters of the 2HDM and thus the mass spectrum of these models are different. 

Flavor physics data also puts the lower limit $m_{H^\pm}>480$ GeV on the charged Higgs mass in the context of the types \RN{2} and \RN{3}~\cite{Misiak}. This constraint results from the dependence of the charged Higgs-quark coupling (corresponding to many flavor observables) on $\tan\beta$ factor in these types. However, the corresponding couplings in types \RN{1} and \RN{4} depend on $\cot\beta$ instead of $\tan\beta$ and thus the behaviour of these types is different from the behaviour of the types \RN{2} and \RN{3}. Therefore, the charged Higgs mass limit in types \RN{1} and \RN{4} are so soft and the selected benchmark points are safe.

Table \ref{signalcrosssection} shows the signal cross section corresponding to different benchmark points and evidently
\begin {table}[h]
\begin{center}
\begin{tabular}{c?c?c?c?c?}
& \multicolumn{4}{ c? }{\cellcolor{blizzardblue}{Benchmark point}} \\ 
& \multicolumn{1}{ c? }{\cellcolor{blizzardblue}BP1} & \cellcolor{blizzardblue}BP2 & \cellcolor{blizzardblue}BP3 & \cellcolor{blizzardblue}BP4 \\ \Xhline{2\arrayrulewidth}
\multicolumn{1}{ c?  }{\cellcolor{blizzardblue}Signal cross section [fb]} & 9.0 & 7.0 & 5.0 & 3.0 \\ \Xhline{2\arrayrulewidth}
\end{tabular}
\caption {Signal cross section corresponding to different benchmark points. \label{signalcrosssection}}
\end{center}
\end {table}
indicates that as the Higgs mass $m_H$ increases the cross section decreases. Although the small cross section in the region of heavier Higgs masses gives rise to a small number of signal events, the narrowness of the tail of the background final resulting distribution (in that mass region) partially compensates for the smallness of the number of signal events and thus to a considerable extent makes searching for heavier Higgs bosons possible. 

Considering the nature of the signal process, the most important background processes include top quark pair production, $W^\pm$ gauge boson pair production, $Z$ gauge boson pair production and finally $Z/\gamma$ production. Table \ref{bgcrosssection} shows the corresponding cross sections which are obtained using PYTHIA 8.2.15~\cite{pythia8.2}.
\begin {table}[h]
\begin{center}
\begin{tabular}{c?c?c?c?c?}
& \multicolumn{4}{ c? }{\cellcolor{blizzardblue}Background process} \\ 
& \multicolumn{1}{ c? }{\cellcolor{blizzardblue}$t\bar{t}$} & \cellcolor{blizzardblue}$WW$ & \cellcolor{blizzardblue}$ZZ$ & \cellcolor{blizzardblue}$Z/\gamma$ \\ \Xhline{2\arrayrulewidth}
\multicolumn{1}{ c?  }{\cellcolor{blizzardblue}Cross section [fb]} & 211.1 & 3163 & 234.7 & 4335 \\ \Xhline{2\arrayrulewidth}
\end{tabular}
\caption {Background cross sections. \label{bgcrosssection}}
\end{center}
\end {table}

\section{Event analysis, selection efficiencies and reconstructed masses} 
The generation of signal events is done in two steps. First, model parameters including Higgs bosons masses and their decay branching ratios are generated in SLHA (SUSY Les Houches Accord) format using 2HDMC 1.6.3 package~\cite{2hdmc1,2hdmc2}. The output files are passed to PYTHIA 8.2.15~\cite{pythia8.2} for event generation and further processing including multi-particle interactions, decays, final state showering, etc. The signal generation is performed for each benchmark point independently. The background event generation is also performed using PYTHIA 8.2.15 for all of the four background processes. 

The jet reconstruction step is performed using FASTJET 3.1.0~\cite{fastjet1,fastjet2} which includes a variety of sequential recombination clustering algorithms. According to the nature of the signal process, the anti-$k_t$ algorithm~\cite{antikt} is chosen as the jet reconstruction algorithm and is expected to give reasonable results. The algorithm uses the standard jet cone size $\Delta R=\sqrt{(\Delta\eta)^2+(\Delta\phi)^2}=0.4$, where $\eta=-\textnormal{ln}\tan(\theta/2)$ and $\phi$ and $\theta$ are the azimuthal and polar angles with respect to the beam axis respectively. 

Jet energy smearing is also applied to jets according to energy resolution $\sigma/E=3.5\%$~\cite{cliccdr}. All jets distinguished by the anti-$k_t$ algorithm are then required to pass the condition $p_{T}\geq10$ GeV, which sets a lower limit on their transverse momenta. Another kinematic limit applied to the resulting jets is defined by the condition $\eta\leq2$ to select only central jets. 

The transverse momentum threshold for muons is set to $p_{T}>5$ GeV. Since the muon production probability in Higgs decays is low, the applied threshold here is lower than jets to compensate the very small branching ratio of Higgs decay to muons ($BR_{H\rightarrow\mu\mu}=0.0035$).

Apart from the di-muon signature of the signal process, which is crucial to the present analysis, the di-taus produced via the decay channel $H \rightarrow \tau \tau$ play a significant role in distinguishing the signal events. The corresponding branching ratio of 0.99 which indicates the dominance of this decay channel over others makes this role even more remarkable. This fact requires the analysis to be equipped with a suitable tau-tagging algorithm by which the tau leptons can be well distinguished. 

Utilizing the single charged pion signature of the tau decay modes $\tau \rightarrow \pi^+ \nu_{\tau}$ and $\tau \rightarrow \pi^+ \pi^0 \pi^0 \nu_{\tau}$, the tau-tagging algorithm first searches for the hottest charged particle in the vicinity of the jet center (defined by $\Delta R<0.1$) and identifies the hottest charged particle as the charged pion $\pi^+$ candidate if it satisfies the transverse momentum condition $p_{T}>10$ GeV. The narrowness of the tau decay can be used as another feature by which the tau jets can be well identified. Because of the approximate collinearity of the tau decay products, the algorithm performs a search in the immediate vicinity of the charged pion candidate (called as the signal cone defined by $\Delta R<0.07$). The number of all found particles in the signal cone is required to be 1 or 3 according to the mentioned tau decay modes. To take full advantage of the narrowness of tau jets, another restriction is applied by requiring that there must be no particle with $p_{T}>1$ GeV in an annulus around the charged pion candidate defined by $0.07\leq\Delta R\leq0.4$. Any jet satisfying the mentioned criteria is finally identified as a tau jet.

To achieve a better understanding of the behavior that the assumed theoretical model, the four hypothesized benchmark points are simulated and tested independently by identical analyses in the present study. Having mentioned the priliminaries, the analysis procedure is discussed in what follows. 

\begin{figure}[h]
  \centering
  \includegraphics[width=.7\textwidth]{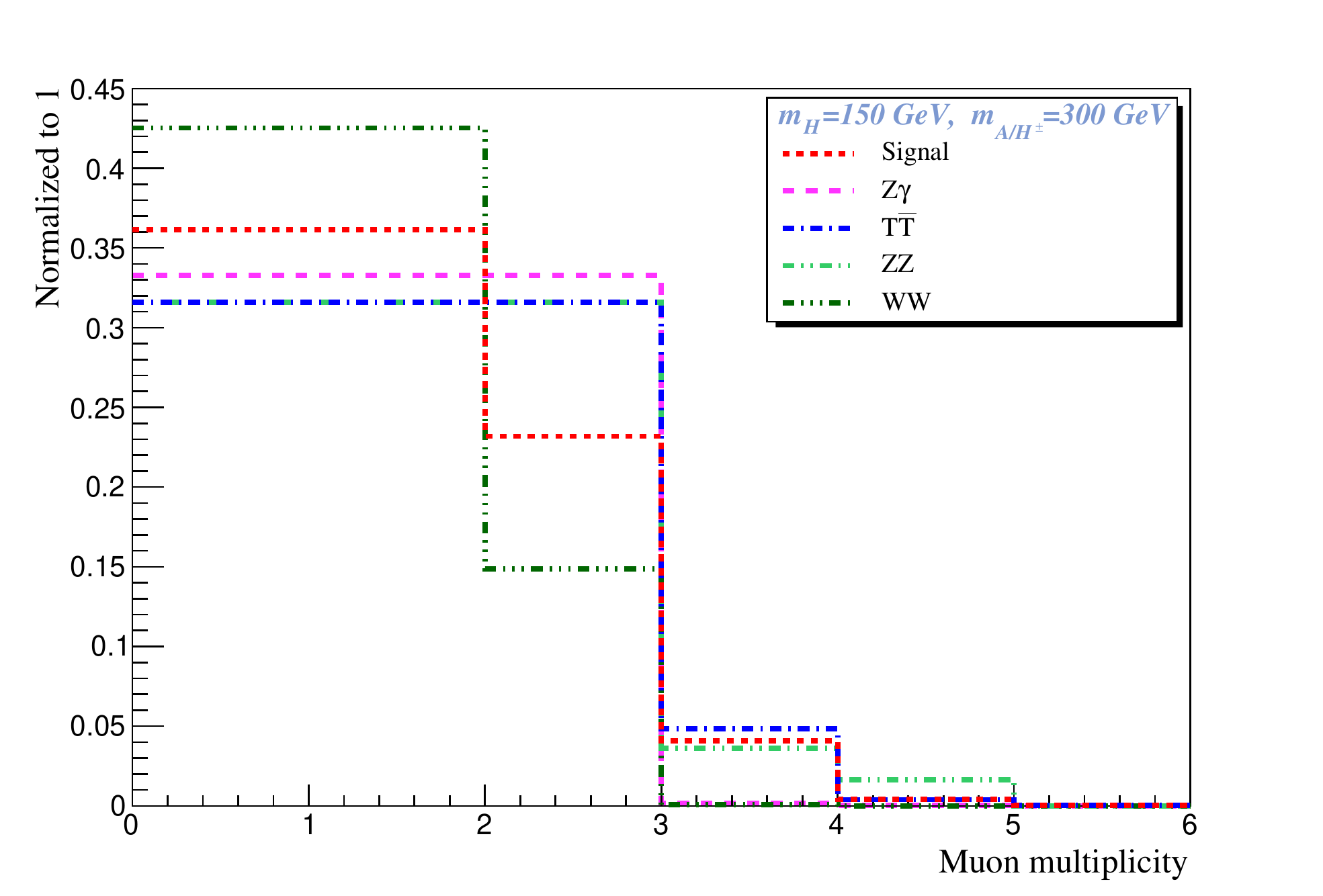}
  \caption{Muon multiplicity obtained using generator level information. \label{muonsmul}}
\end{figure}

The analysis begins by identifying muon leptons using the information in generator level. The identified muons are required to meet a transverse momentum threshold condition by applying the cut
\begin{equation}
\bm{p_T}>5\ GeV.  
\label{leptonptcut}
\end{equation}
Counting the passed muons a muon multiplicity distribution is obtained for each process. The resulting distributions of the signal and different background processes corresponding to the benchmark point 1 (BP1) are shown in figure \ref{muonsmul}.
Di-muons will be used to reconstruct the Higgs boson $H$ mass at the end of the analysis and thus the condition
\begin{equation}
No.\ of\ \textbf{\emph{muons}}\geq2
\label{no.ofmuonscut}
\end{equation}
is applied to make sure the needed di-muons exist in the events.

\begin{figure}[h]
  \centering
  \includegraphics[width=.7\textwidth]{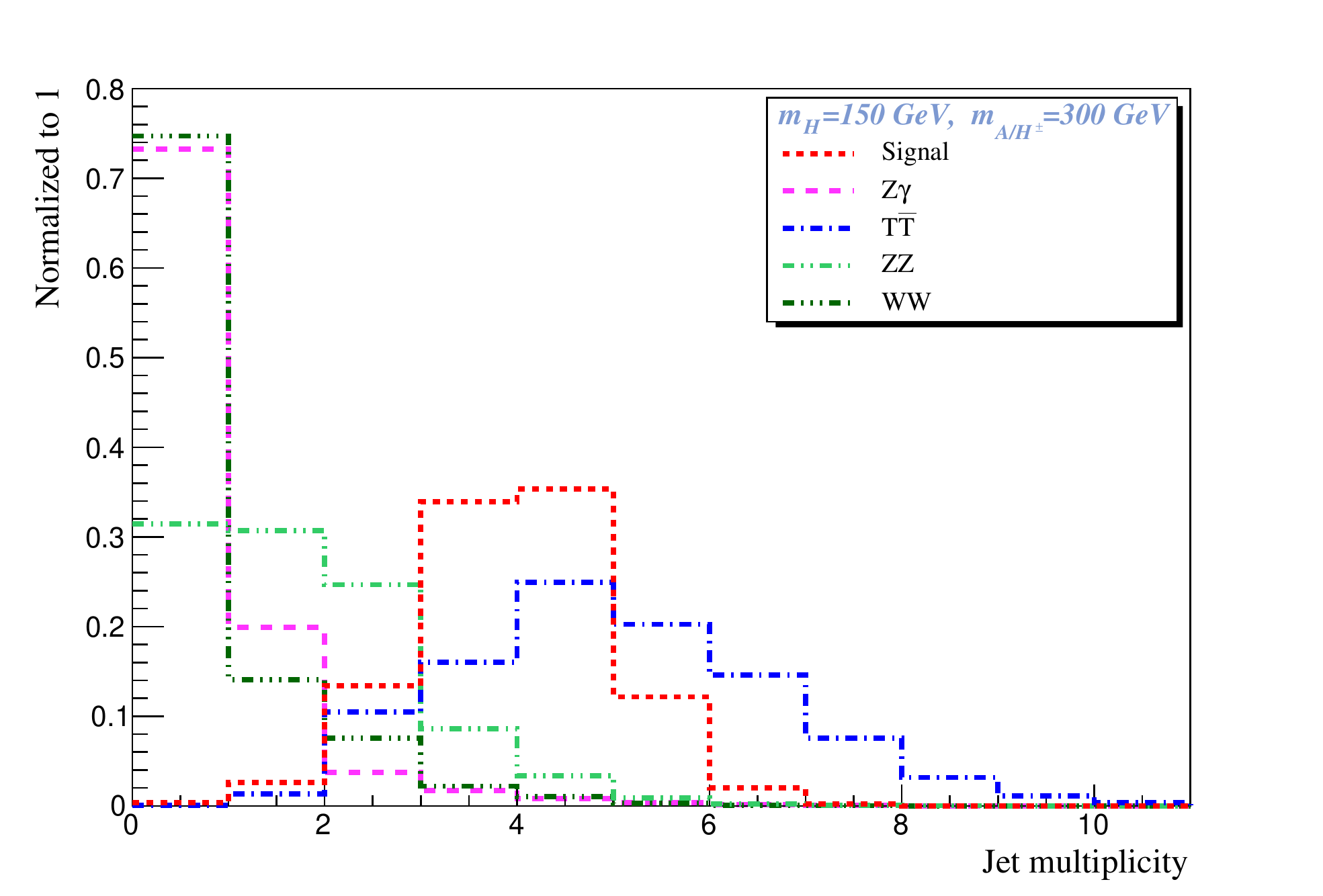}
  \caption{Jet multiplicity obtained using Anti-$k_t$ algorithm. \label{jetmul}}
\end{figure}

In the next step, a search for standard jets is performed using the anti-$k_t$ jet reconstruction algorithm and the resulting jets are examined to see if the kinematic criteria 
\begin{equation}
\bm{p_T}>10\ GeV,\ \ \  \bm{\eta}<2,
\label{jetptcut}
\end{equation}
which suit our needs are satisfied or not. Counting the jets passing these criteria the plot of figure \ref{jetmul} is obtained for the multiplicity of jets in signal and background events.
As seen in figure \ref{jetmul}, for all the background processes except $t\bar{t}$ the jet multiplicity distribution seems to follow a significantly different pattern from the signal pattern. Based on this difference, the cut 
\begin{equation}
No.\ of\ \textbf{\emph{jets}}\geq4
\label{no.ofjetscut}
\end{equation}
is defined and applied to refine the selected events. 

\begin{figure}[h]
  \centering
  \includegraphics[width=.7\textwidth]{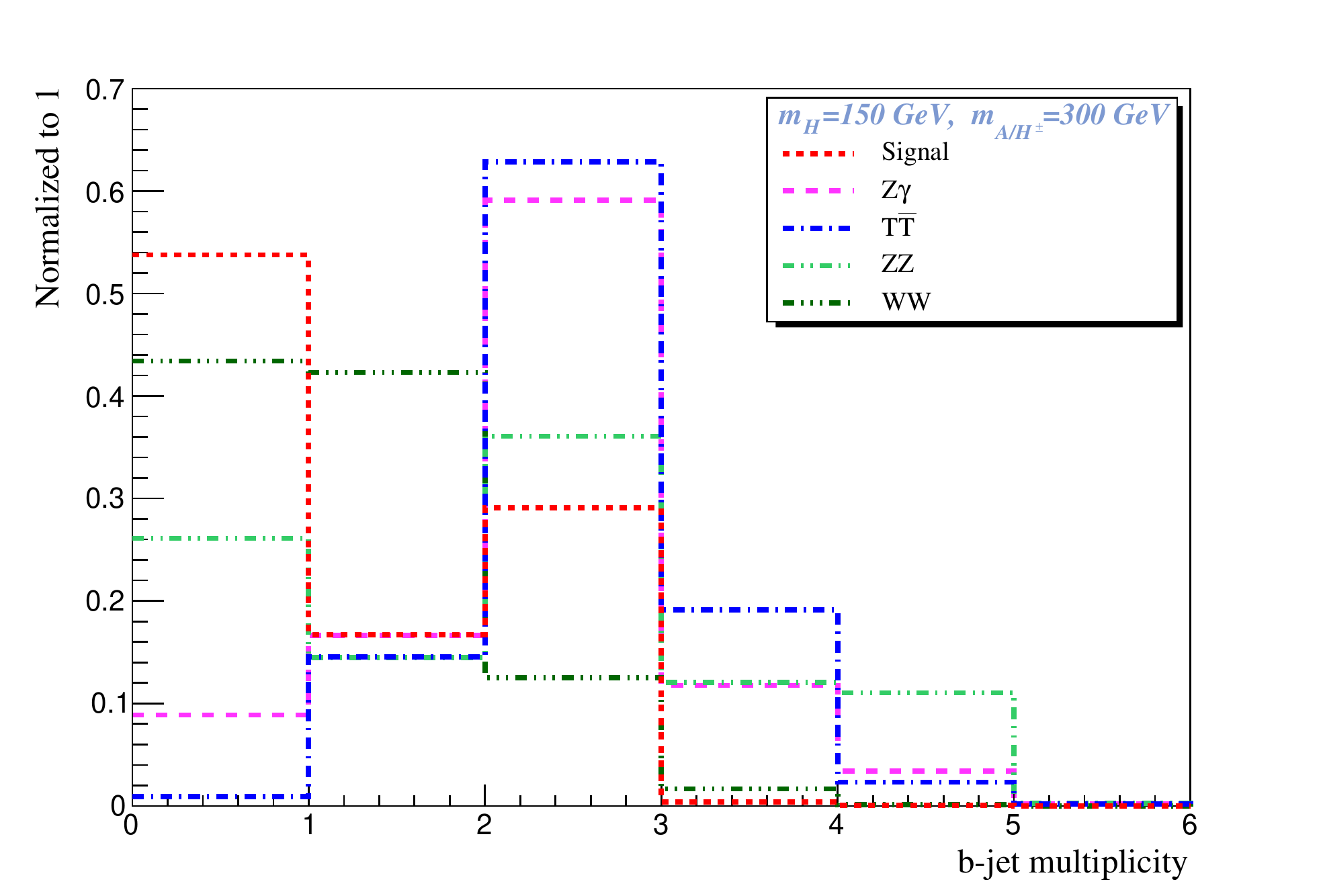}
  \caption{b-jet multiplicity obtained by b-tagging algorithm. \label{bjetmul}}
\end{figure}

A $b$-tagging algorithm is then applied to jets of the survived events to find the number of $b$-jets included in each event. The used $b$-tagging algorithm performs a search for adjacent $b$ or $c$ quarks using the information in generator level for each selected jet. A jet is identified as a $b$-jet with 60\% (10\%) probability if it is near a $b$ ($c$) quark. Having applied the $b$-tagging algorithm a $b$-jet multiplicity distribution is obtained. Figure \ref{bjetmul} shows the $b$-jet multiplicity distributions.
As seen in figure \ref{bjetmul}, all the background processes except the $WW$ process follow a different pattern from the signal pattern. To take advantage of this contrast the cut
\begin{equation}
No.\ of\ \textbf{\emph{b-jets}}\leq1 
\label{no.ofbjetscut}
\end{equation}
is applied. As it's obvious from figure \ref{bjetmul}, the majority of signal events include no $b$-jet or only one $b$-jet (with less probability). It's due to the fact that the main source of the $b$ quark in the signal process is the $Z$ boson decay to $b\bar{b}$ pair which its branching ratio is relatively small.

\begin{figure}[h]
  \centering
  \includegraphics[width=.7\textwidth]{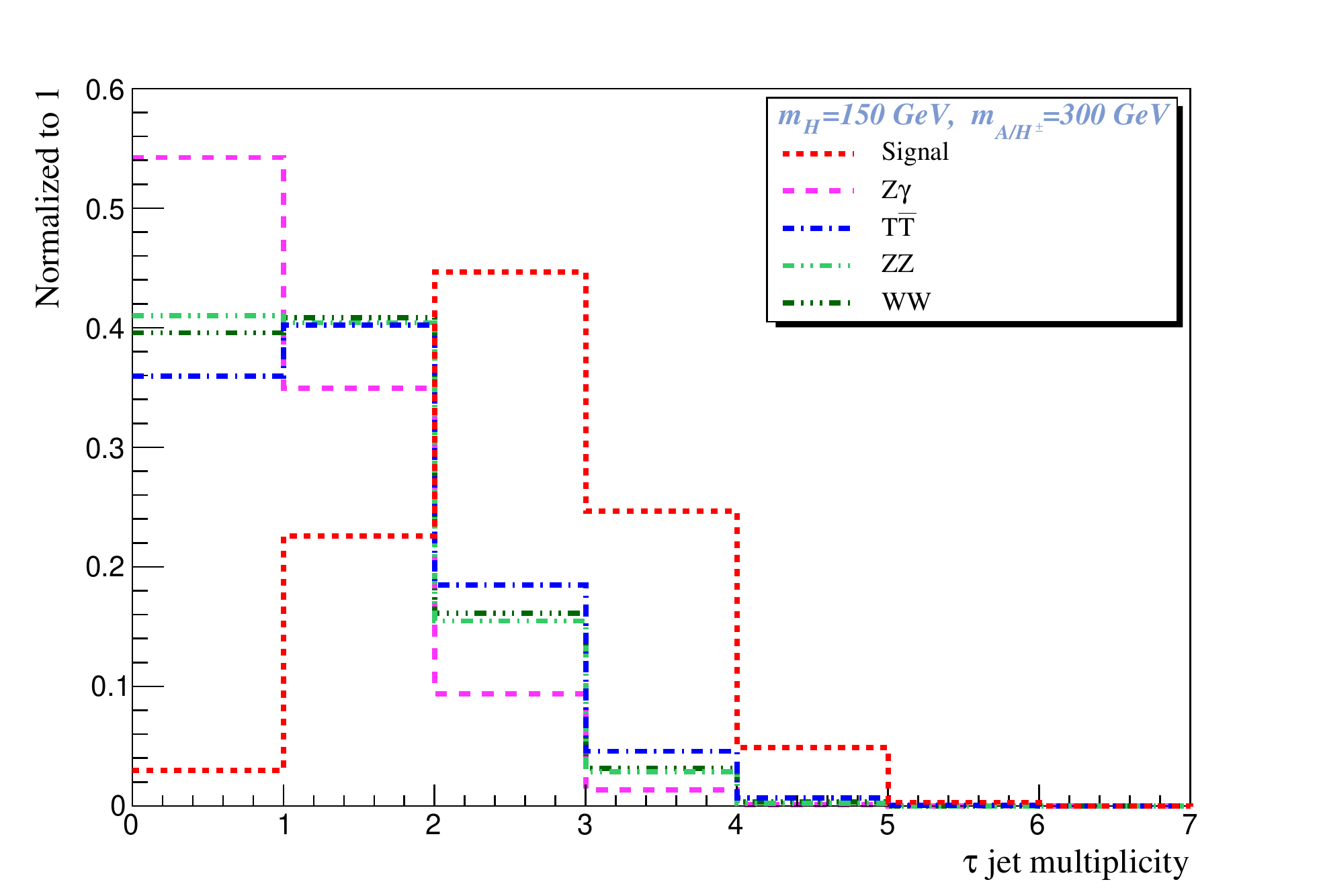}
  \caption{$\tau$-jet multiplicity obtained using the $\tau$-tagging algorithm. \label{taujetmul}}
\end{figure}

Considering the signal process, the jets in the signal events originate mainly from the products of the decay channels $H \rightarrow \tau \tau$ and $Z\rightarrow q\bar{q}$. Hence the $\tau$-tagging algorithm is now applied to jets to distinguish $\tau$-jets from the others. Having performed the $\tau$-tagging algorithm, the plot of figure \ref{taujetmul} is obtained which shows the $\tau$-jet multiplicity of the various processes. 
According to figure \ref{taujetmul}, the average number of $\tau$-jets in the signal events is greater than that of the background events. This can be explained by the relatively high branching ratio of 0.99 for the Higgs boson $H$ decay to a tauon pair. The distributions shown in figure \ref{taujetmul} suggest applying the cut 
\begin{equation}
No.\ of\ \bm{\tau}\textbf{\emph{-jets}}\geq2     
\label{no.oftaujetscut}
\end{equation}
to the selected events in the previous step. 

\begin{figure}[h]
  \centering
  \includegraphics[width=.7\textwidth]{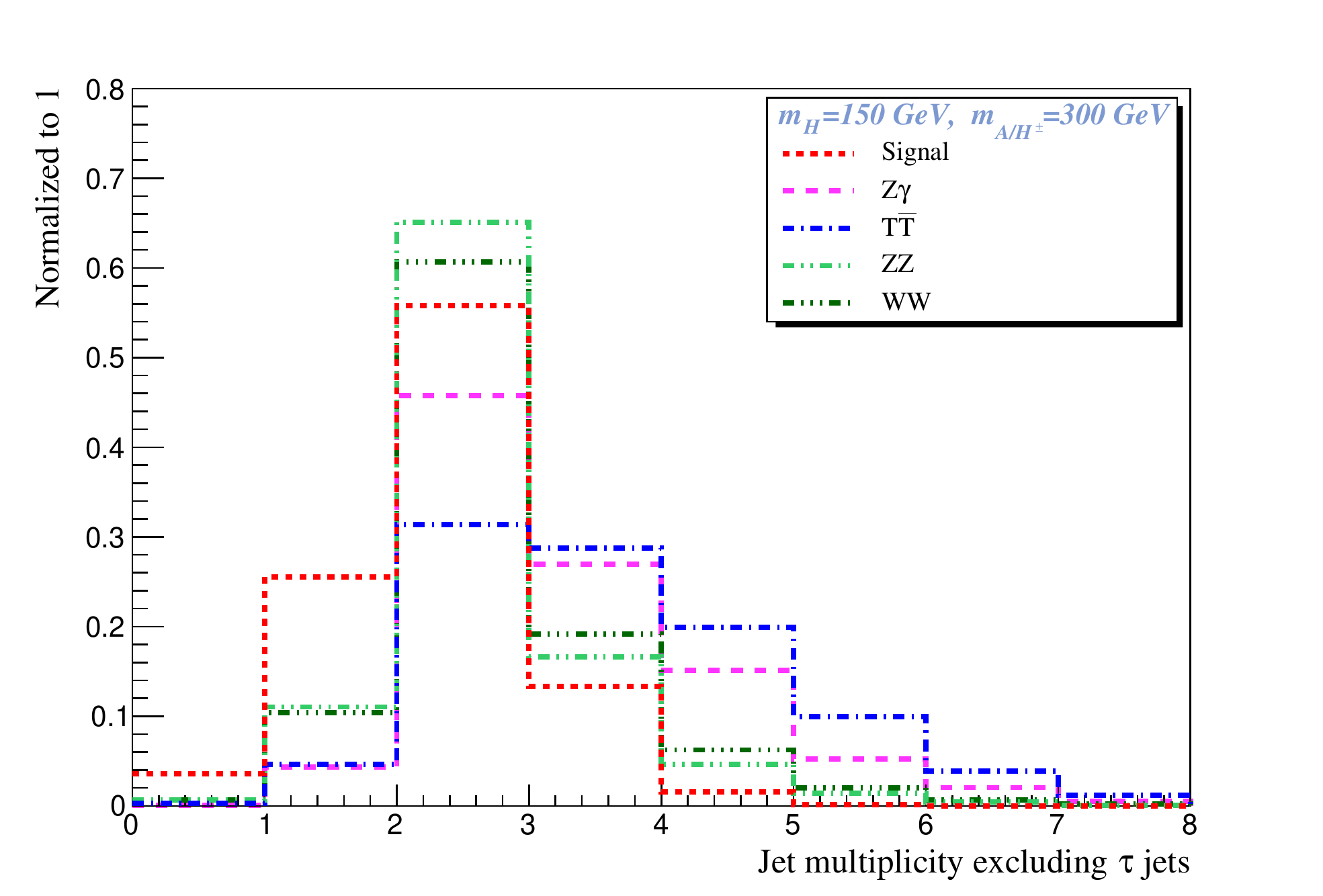}
  \caption{Jet multiplicity after applying cut \ref{no.oftaujetscut} and excluding $\tau$-jets. \label{jetsexcludingtaujets}}
\end{figure}

\begin{figure}[h]
  \centering
  \includegraphics[width=.7\textwidth]{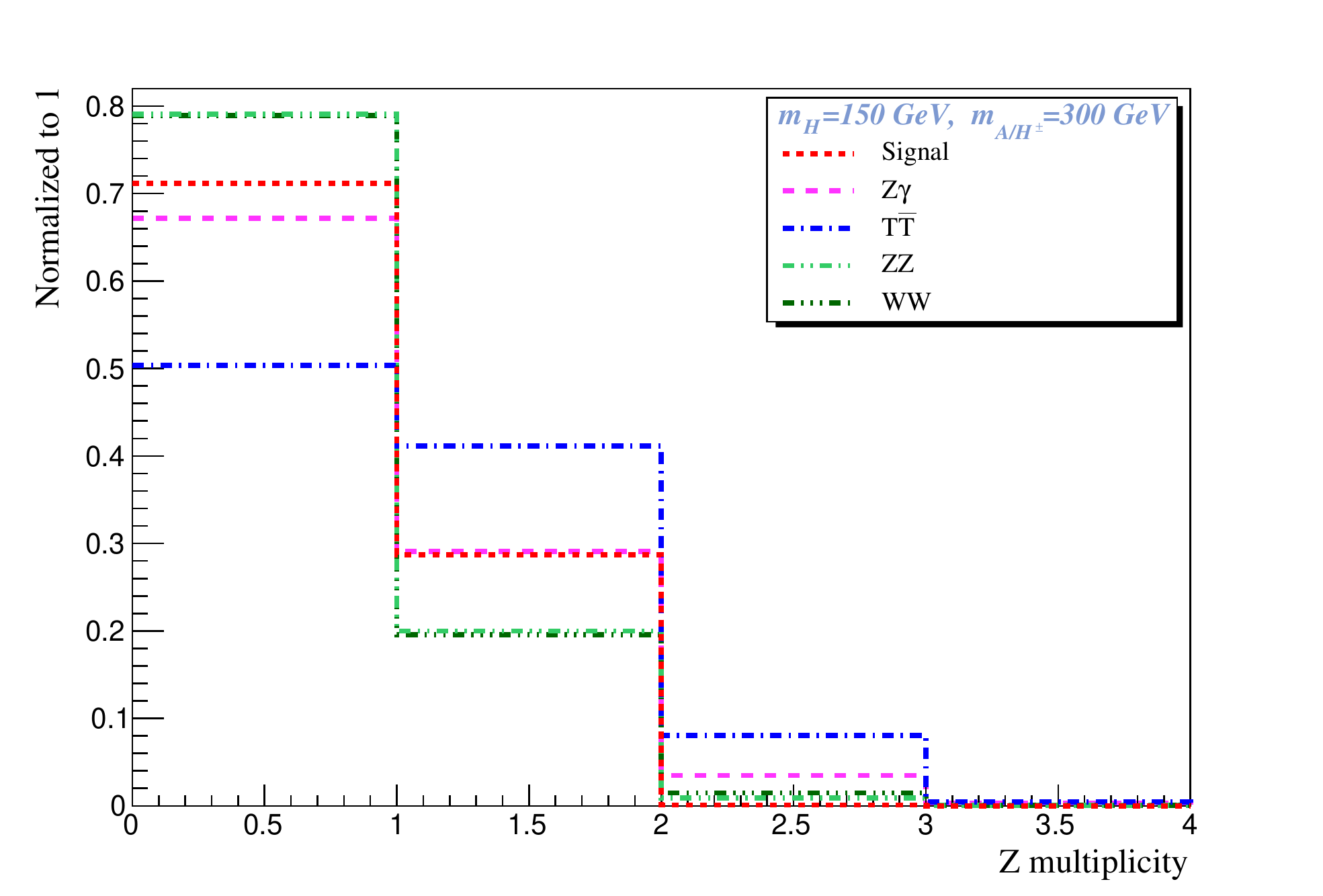}
  \caption{$Z$ multiplicity obtained by applying mass window cut \ref{zmasswindowcut}. \label{zjetmul}}
\end{figure}

Figure \ref{jetsexcludingtaujets} shows the jet multiplicity after applying the cut \ref{no.oftaujetscut} and excluding $\tau$-jets. 
The remaining jets when $\tau$-jets are excluded are candidates for decay products of the $Z$ boson. Therefore, the invariant mass of each possible pair of them is calculated and tested by the mass window cut 
\begin{equation}
70.0< \bm{M_{inv}} < 110.0
\label{zmasswindowcut}
\end{equation}
to assess its origin. Having tested all the possible pairs, the $Z$ multiplicity is obtained as shown in figure \ref{zjetmul}.
Since the signal and background distributions of figures \ref{jetsexcludingtaujets} and \ref{zjetmul} follow similar patterns, no cut is applied on the number of $Z$ bosons.

Applying all cuts to signal and background events, relative and total efficiencies are obtained as shown in table \ref{signalefftab} for signal events and table \ref{backgroundefftab} for different background processes.
\begin{table}[h]
\normalsize
\fontsize{11}{7.2} 
    \begin{center}
         \begin{tabular}{ >{\centering\arraybackslash}m{1.9in} ? >{\centering\arraybackslash}m{.6in} ? >{\centering\arraybackslash}m{.6in} ?>{\centering\arraybackslash}m{.6in} ? >{\centering\arraybackslash}m{.6in} ? >{\centering\arraybackslash}m{.6in} ?} 
& \cellcolor{blizzardblue}{BP 1} & \cellcolor{blizzardblue}{BP 2} & \cellcolor{blizzardblue}{BP 3} & \cellcolor{blizzardblue}{BP 4} \parbox{0pt}{\rule{0pt}{1ex+\baselineskip}}\\ \Xhline{6\arrayrulewidth}
    \cellcolor{blizzardblue}{Muons number} & 0.152 & 0.156 & 0.159 & 0.161  \parbox{0pt}{\rule{0pt}{1ex+\baselineskip}}\\ \Xhline{2\arrayrulewidth}
    \cellcolor{blizzardblue}{Jets number} & 0.497 & 0.533 & 0.554 & 0.561  \parbox{0pt}{\rule{0pt}{1ex+\baselineskip}}\\ \Xhline{2\arrayrulewidth}
    \cellcolor{blizzardblue}{$b$-tagging} & 0.705 & 0.673 & 0.645 & 0.636  \parbox{0pt}{\rule{0pt}{1ex+\baselineskip}}\\ \Xhline{2\arrayrulewidth}
    \cellcolor{blizzardblue}{$\tau$-tagging} & 0.745 & 0.752 & 0.763 & 0.775  \parbox{0pt}{\rule{0pt}{1ex+\baselineskip}}\\ \Xhline{2\arrayrulewidth}
   \cellcolor{blizzardblue}{Total eff.} & 0.040 & 0.042 & 0.043 & 0.045  \parbox{0pt}{\rule{0pt}{1ex+\baselineskip}}\\ \Xhline{2\arrayrulewidth}
  \end{tabular}
\caption{Signal selection efficiencies assuming different benchmark points. \label{signalefftab}}
  \end{center}
\end{table}
\begin{table}[h]
\normalsize
\fontsize{11}{7.2} 
    \begin{center}
         \begin{tabular}{ >{\centering\arraybackslash}m{1.9in} ? >{\centering\arraybackslash}m{.6in} ? >{\centering\arraybackslash}m{.6in} ?>{\centering\arraybackslash}m{.6in} ? >{\centering\arraybackslash}m{.6in} ? >{\centering\arraybackslash}m{.6in} ?}
& \cellcolor{blizzardblue}{$t\bar{t}$} & \cellcolor{blizzardblue}{$WW$} & \cellcolor{blizzardblue}{$ZZ$} & \cellcolor{blizzardblue}{$Z/\gamma$ } \parbox{0pt}{\rule{0pt}{1ex+\baselineskip}}\\ \Xhline{6\arrayrulewidth}
    \cellcolor{blizzardblue}{Muons number} & 0.106 & 0.019 & 0.099 & 0.066  \parbox{0pt}{\rule{0pt}{1ex+\baselineskip}}\\ \Xhline{2\arrayrulewidth}
    \cellcolor{blizzardblue}{Jets number} & 0.716 & 0.014 & 0.046 & 0.012  \parbox{0pt}{\rule{0pt}{1ex+\baselineskip}}\\ \Xhline{2\arrayrulewidth}
    \cellcolor{blizzardblue}{$b$-tagging} & 0.155 & 0.857 & 0.406 & 0.255  \parbox{0pt}{\rule{0pt}{1ex+\baselineskip}}\\ \Xhline{2\arrayrulewidth}
    \cellcolor{blizzardblue}{$\tau$-tagging} & 0.238 & 0.196 & 0.186 & 0.108  \parbox{0pt}{\rule{0pt}{1ex+\baselineskip}}\\ \Xhline{2\arrayrulewidth}
    \cellcolor{blizzardblue}{Total eff.} & 0.0028 & 0.00005 & 0.0003 & 0.00002  \parbox{0pt}{\rule{0pt}{1ex+\baselineskip}}\\ \Xhline{2\arrayrulewidth}
  \end{tabular}
\caption{Background selection efficiencies. \label{backgroundefftab}}
  \end{center}
\end{table}

\begin{figure}[h]
  \centering
  \includegraphics[width=.7\textwidth]{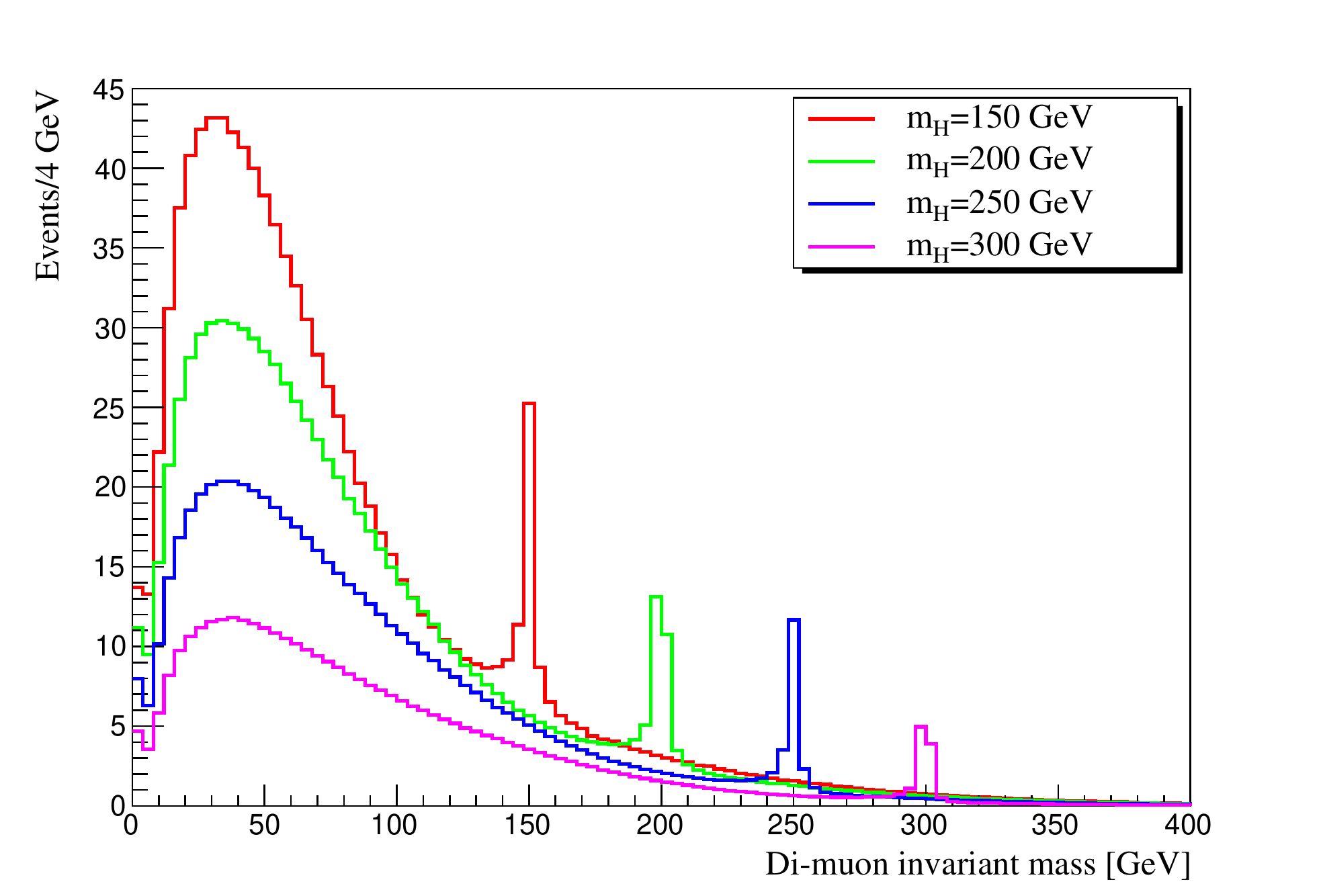}
  \caption{Higgs boson $H$ reconstructed mass in signal events for different assumed benchmark points. $m_H=150, 200, 250$ and $300$ GeV correspond to benchmark points 1, 2, 3 and 4 respectively.\label{HallmassesS}}
\end{figure}

The distinguished di-muons in signal events come from the Higgs boson $H$ and therefore their invariant mass must be in principle equal to Higgs $H$ mass. However, jet energy resolution, mis-identification of jets and also errors in energy and flight directions of particles result in an invariant mass distribution with a peak almost at the generated Higgs boson $H$ mass. Figure \ref{HallmassesS} shows the di-muon invariant mass distribution in signal events for the four assumed benchmark points. The total number of expected events to use for nomalization is obtained from $\sigma\times\epsilon\times L$, where $\sigma$ is the signal cross section (from table \ref{signalcrosssection}), $\epsilon$ is the total efficiency (from table \ref{signalefftab}) and $L$ is the integrated luminosity which is assumed to be 3000 $fb^{-1}$.

\begin{figure}[h]
  \centering
  \includegraphics[width=.7\textwidth]{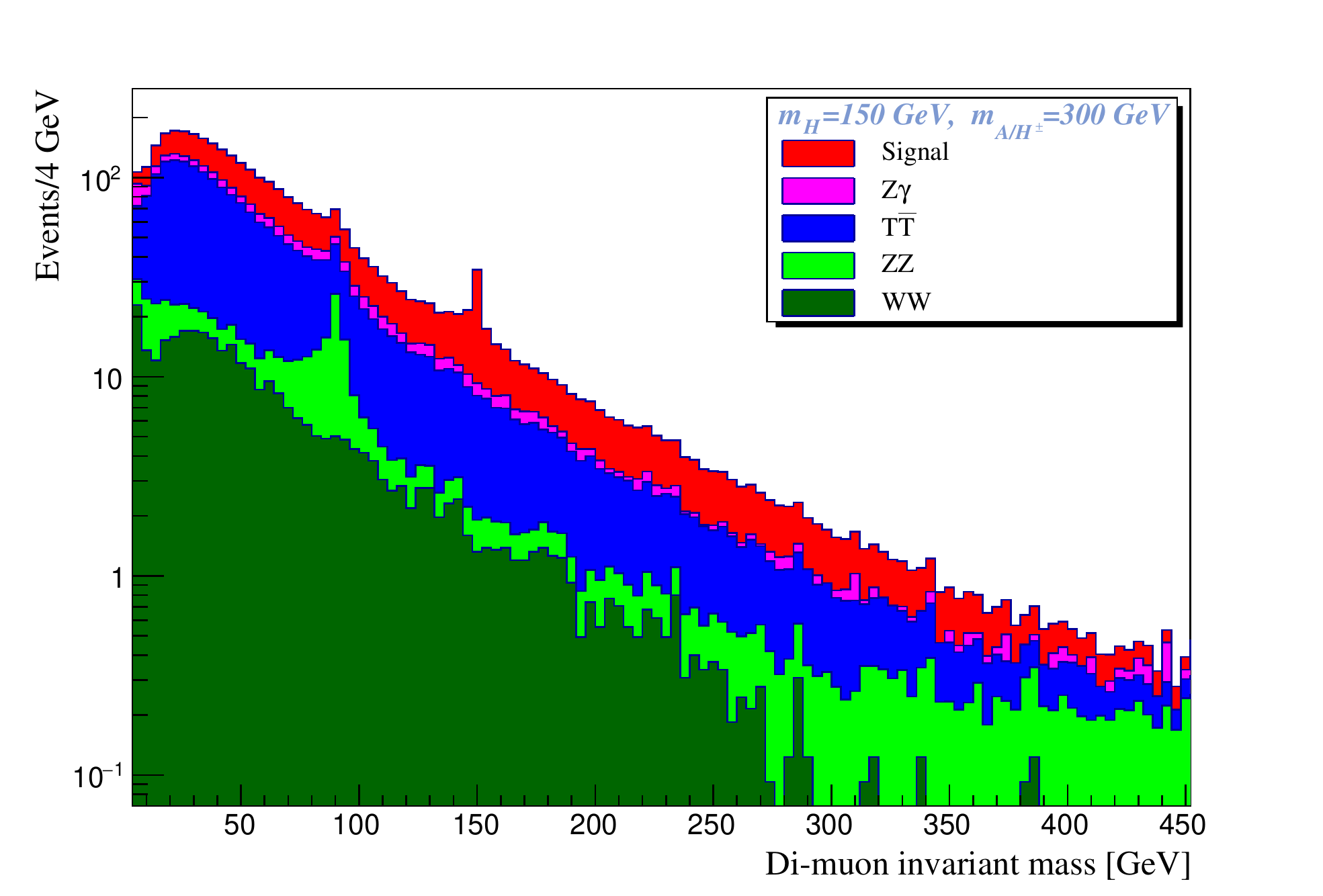}
  \caption{Distribution of signal plus backgrounds assuming BP1. \label{H150}}
\end{figure}

\begin{figure}[h]
  \centering
  \includegraphics[width=.7\textwidth]{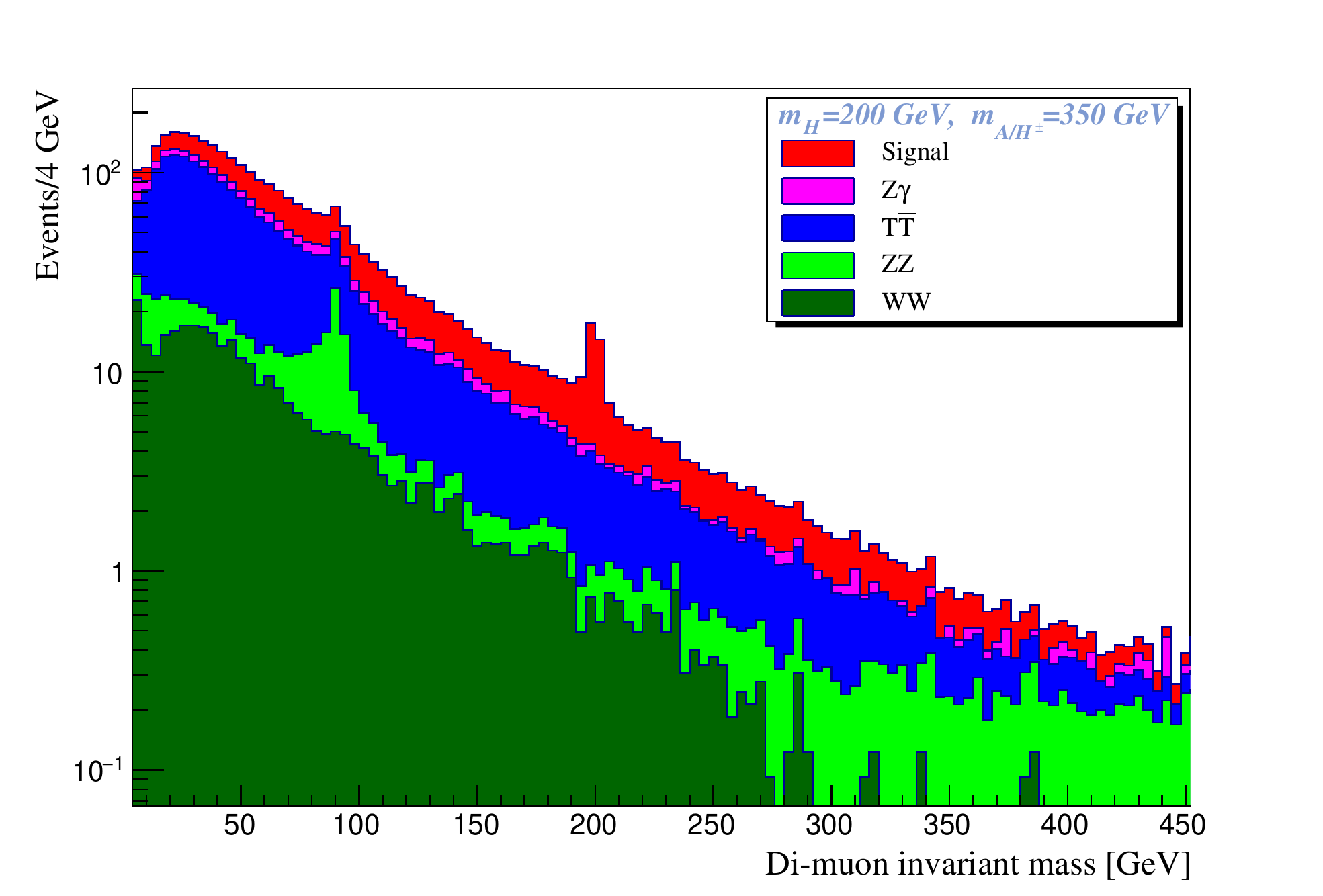}
  \caption{Distribution of signal plus backgrounds assuming BP2. \label{H200}}
\end{figure}

\begin{figure}[h]
  \centering
  \includegraphics[width=.7\textwidth]{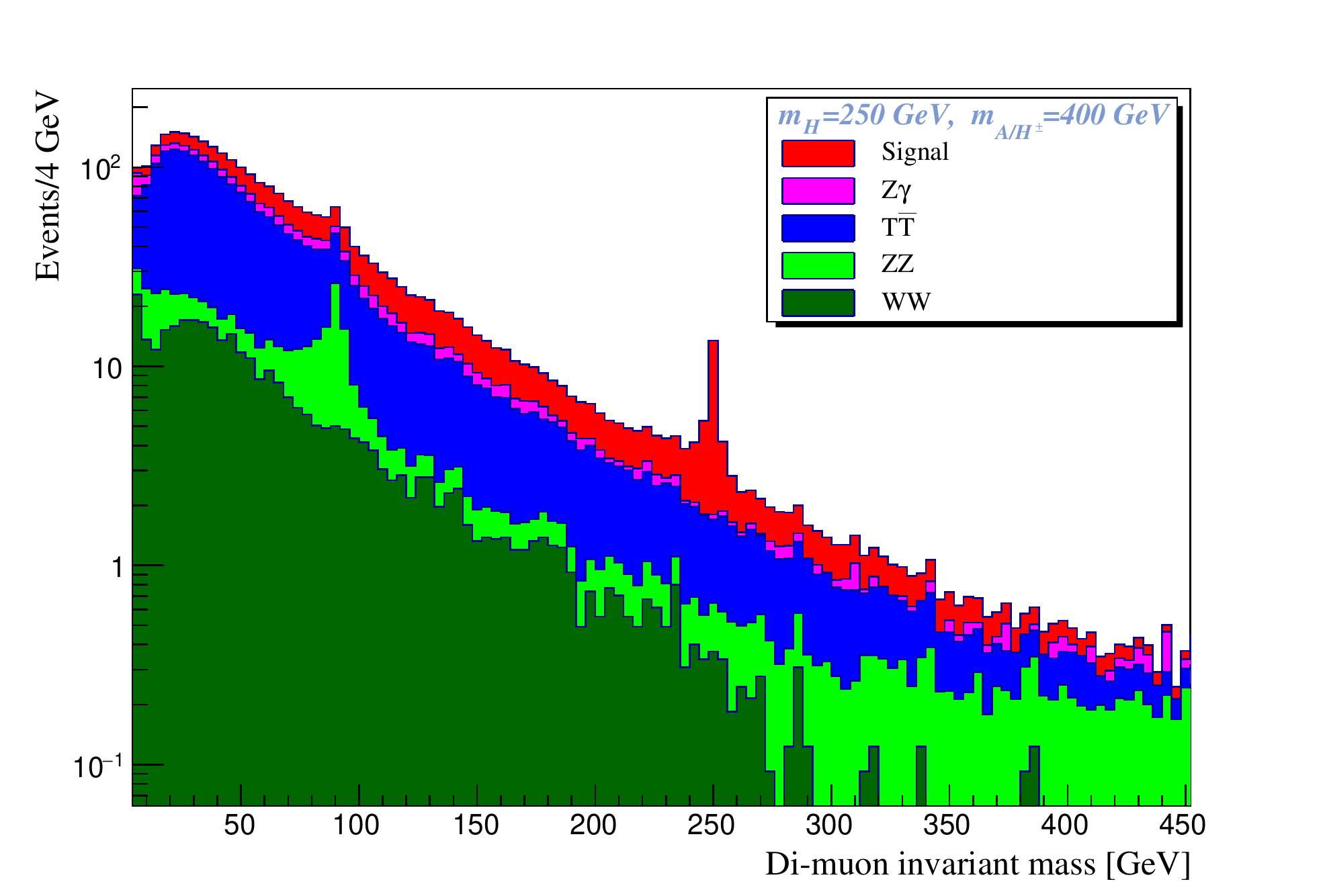}
  \caption{Distribution of signal plus backgrounds assuming BP3. \label{H250}}
\end{figure}

\begin{figure}[h]
  \centering
  \includegraphics[width=.7\textwidth]{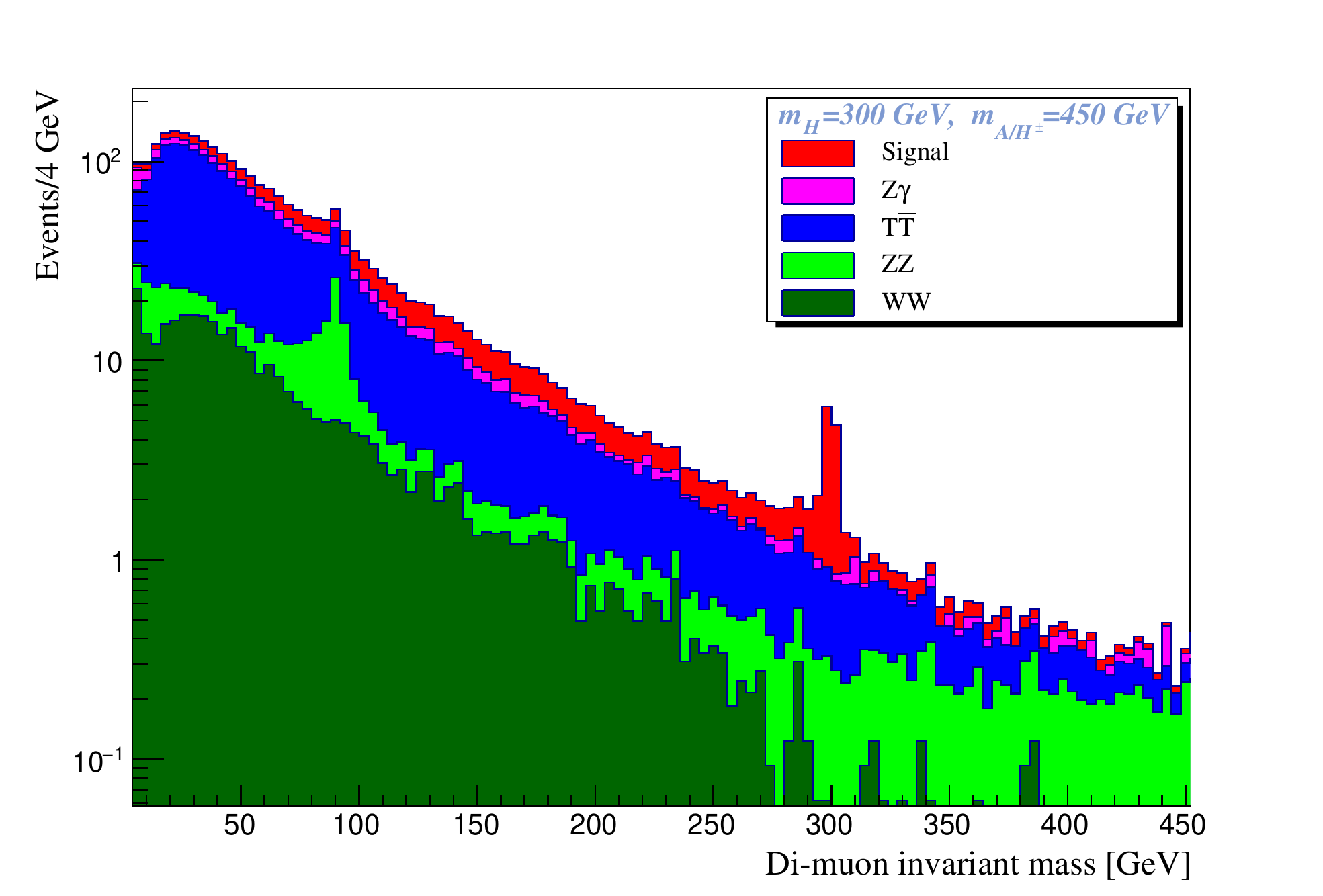}
  \caption{Distribution of signal plus backgrounds assuming BP4. \label{H300}}
\end{figure}

Figures \ref{H150}-\ref{H300} show the signal distribution on top of the background distributions for different benchmark points. As the figures show, the signal events can be seen as relatively small excess of events on top of the background events. This is due the fact that the background processes possess larger cross section than the signal processes. Apart from the apparent peak centered almost at the Higgs generated mass, all the distributions indicate a small peak around $Z$ boson mass which is due to the decaying $Z$ bosons resulting mainly from the $ZZ$ background events. 

\begin{figure}[h]
  \centering
  \includegraphics[width=.7\textwidth]{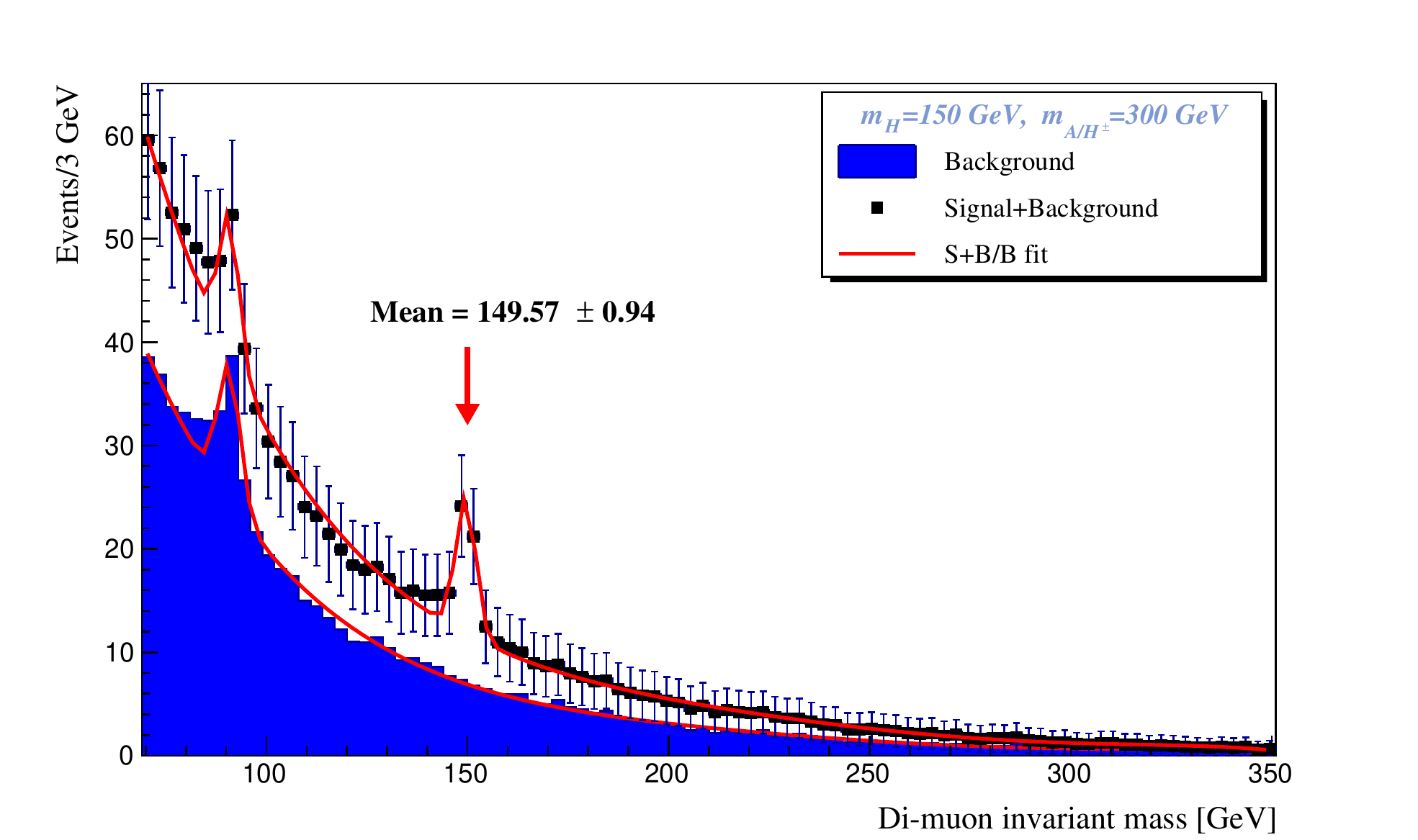}
  \caption{Signal events on top of the total background assuming BP1. Solid curves show the background fit and the fit to signal plus total background. The data with statistical error bars and the mean value of the Gaussian fit function are also shown. \label{150fit}}
\end{figure}

\begin{figure}[h]
  \centering 
  \includegraphics[width=.7\textwidth]{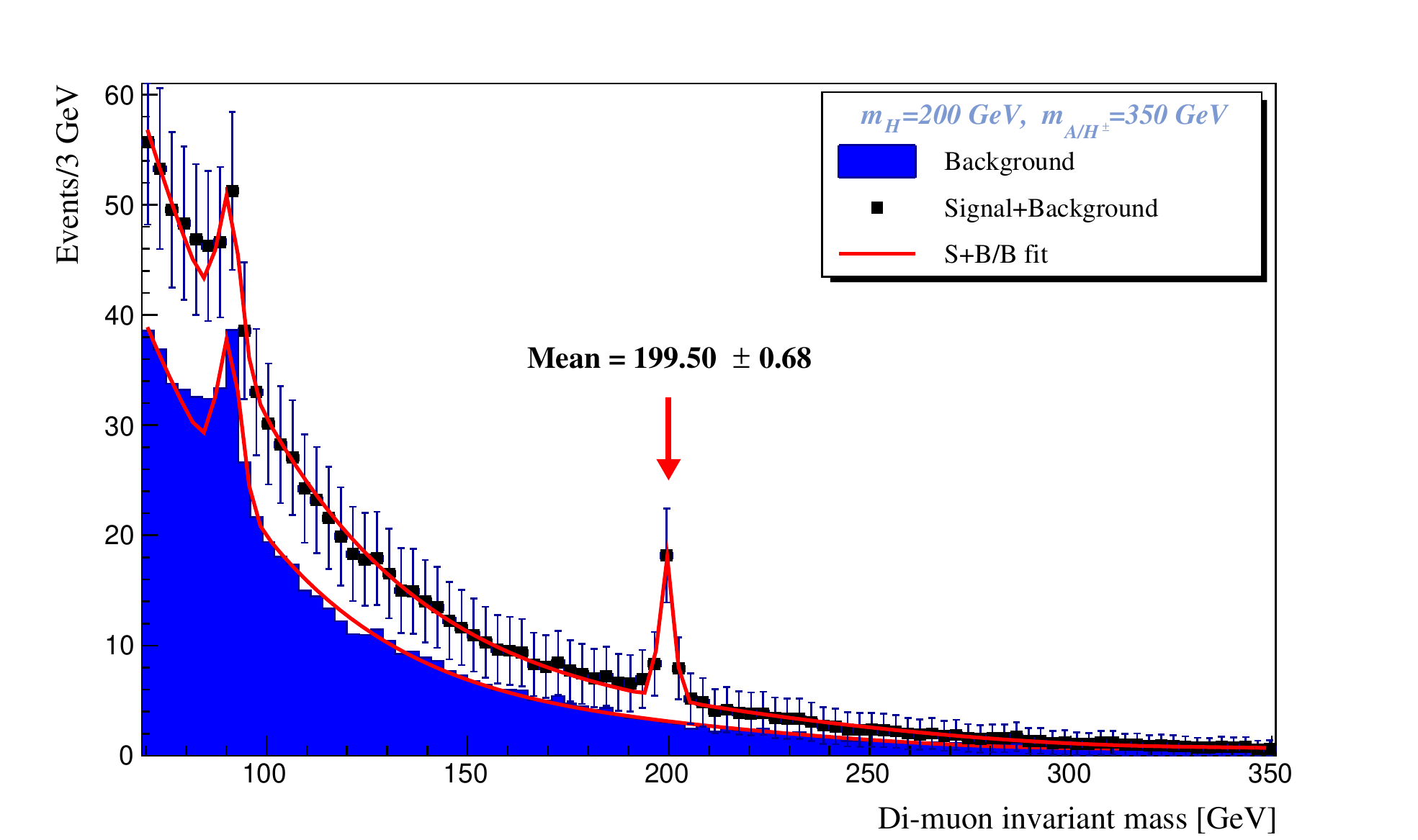}
  \caption{Signal events on top of the total background assuming BP2. Solid curves show the background fit and the fit to signal plus total background. The data with statistical error bars and the mean value of the Gaussian fit function are also shown. \label{200fit}}
\end{figure}

\begin{figure}[h]
  \centering
  \includegraphics[width=.7\textwidth]{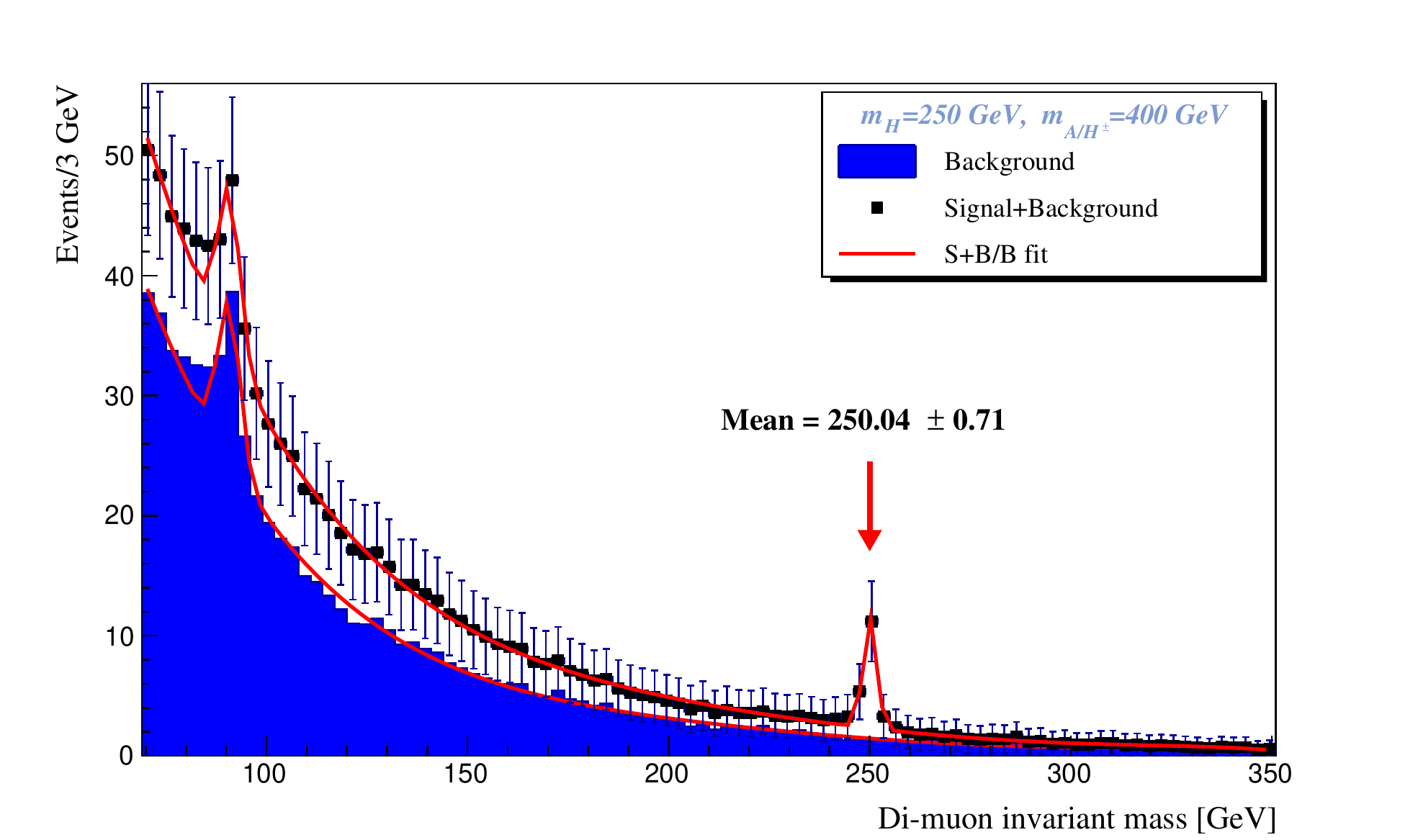}
  \caption{Signal events on top of the total background assuming BP3. Solid curves show the background fit and the fit to signal plus total background. The data with statistical error bars and the mean value of the Gaussian fit function are also shown. \label{250fit}}
\end{figure}

\begin{figure}[h]
  \centering
  \includegraphics[width=.7\textwidth]{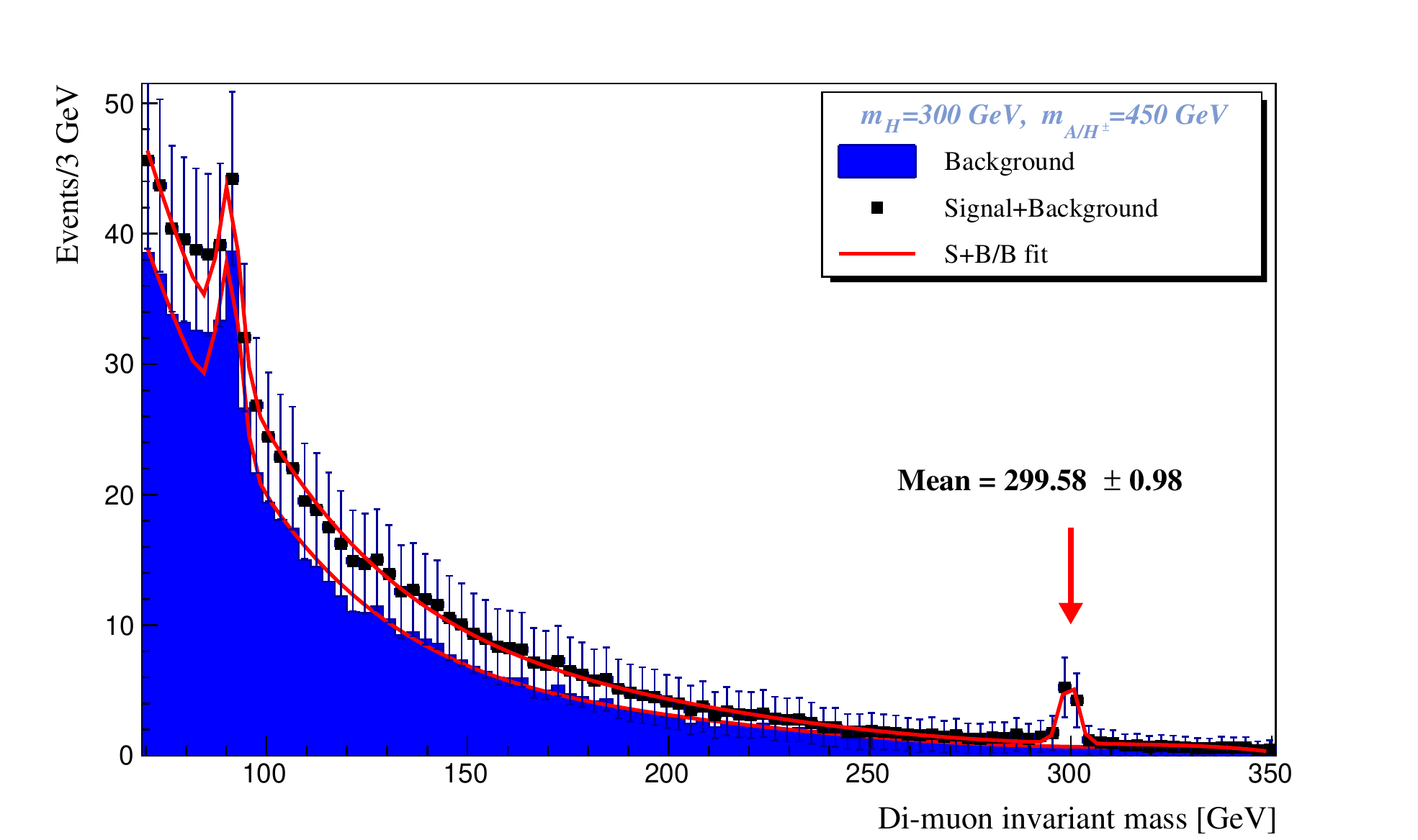}
  \caption{Signal events on top of the total background assuming BP4. Solid curves show the background fit and the fit to signal plus total background. The data with statistical error bars and the mean value of the Gaussian fit function are also shown. \label{300fit}}
\end{figure}

The Higgs candidate mass distributions are now used for the Higgs mass reconstruction. We construct two functions which have the best fit to the signal plus background and background distributions using ROOT 5.34~\cite{root}. The Higgs boson mass is read from the right fit parameter. The fit function corresponding to the signal plus background distribution is a combination of a polynomial function along with two Gaussian functions. The two Gaussian functions cover the Higgs and $Z$ peaks. The fit parameters of one of the Gaussians may be used to determine the Higgs reconstructed mass. For the background distribution, a combination of a polynomial and a Gaussian function is used as the fit function to be fitted to the distribution. Figures \ref{150fit}-\ref{300fit} show the fits results corresponding to the four distributions in figures \ref{H150}-\ref{H300}.   
\begin{figure}[h]
  \centering
  \includegraphics[width=.55\textwidth]{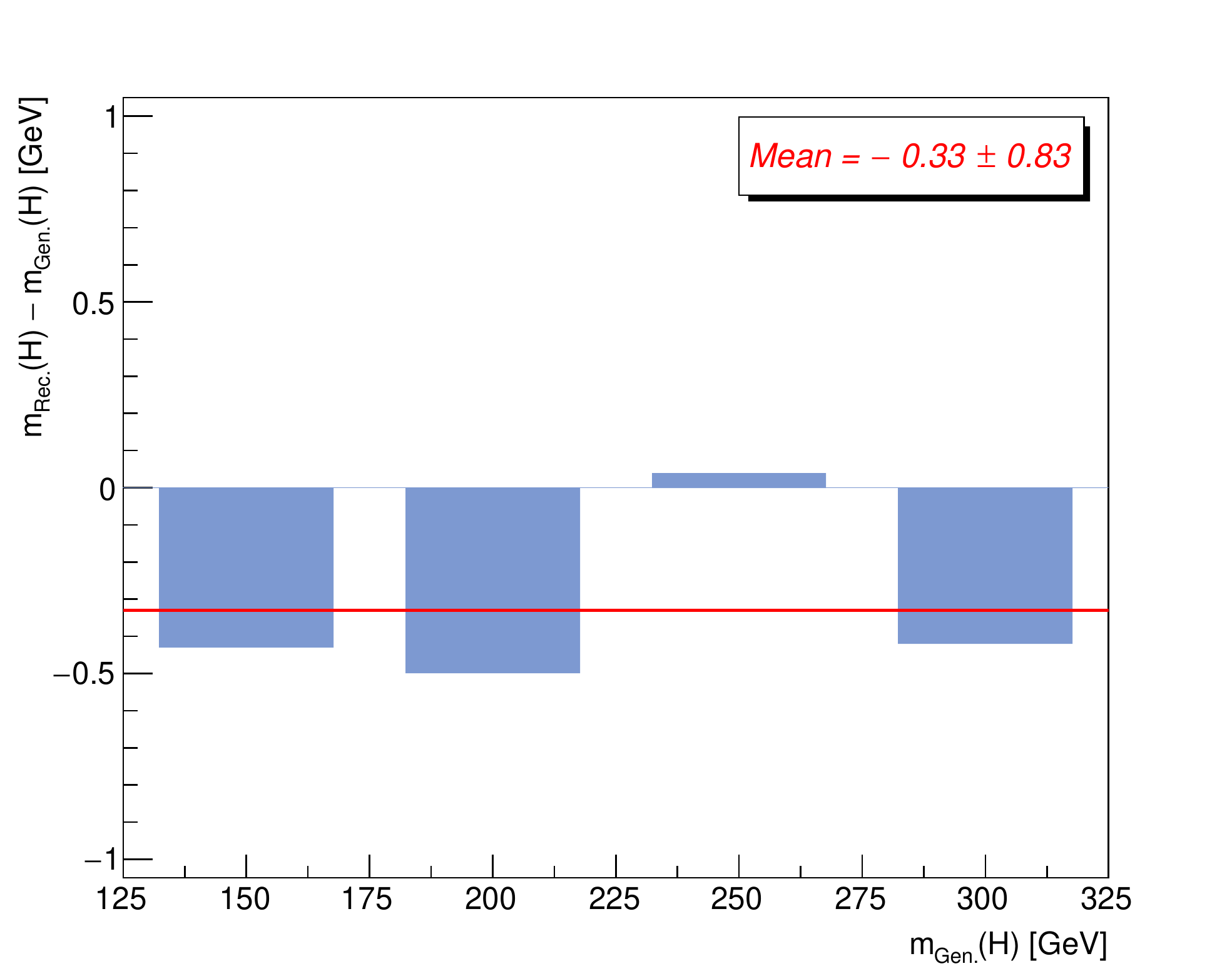}
  \caption{Reconstructed Higgs mass minus generated Higgs mass corresponding to the four assumed benchmark points. \label{recmassMinusGenmass}}
\end{figure}

As seen in figures \ref{150fit}-\ref{300fit}, the signal is well distinguished from the background for all of the assumed benchmark points. The value of the ``mean'' parameter of the Gaussian fit function is close to the generated Higgs mass. However there is small off-set which is shown in figure \ref{recmassMinusGenmass}.  This difference can be due to the jet reconstruction algorithm and uncertainties arising from it. The jet reconstruction parameters can be tuned so that the algorithm works as well as possible. A thorough study of the jet reconstruction results and the generated particles and a comparison between them using MC truth matching tools can clarify the error sources. 

\begin{figure}[h]
  \centering
  \includegraphics[width=.55\textwidth]{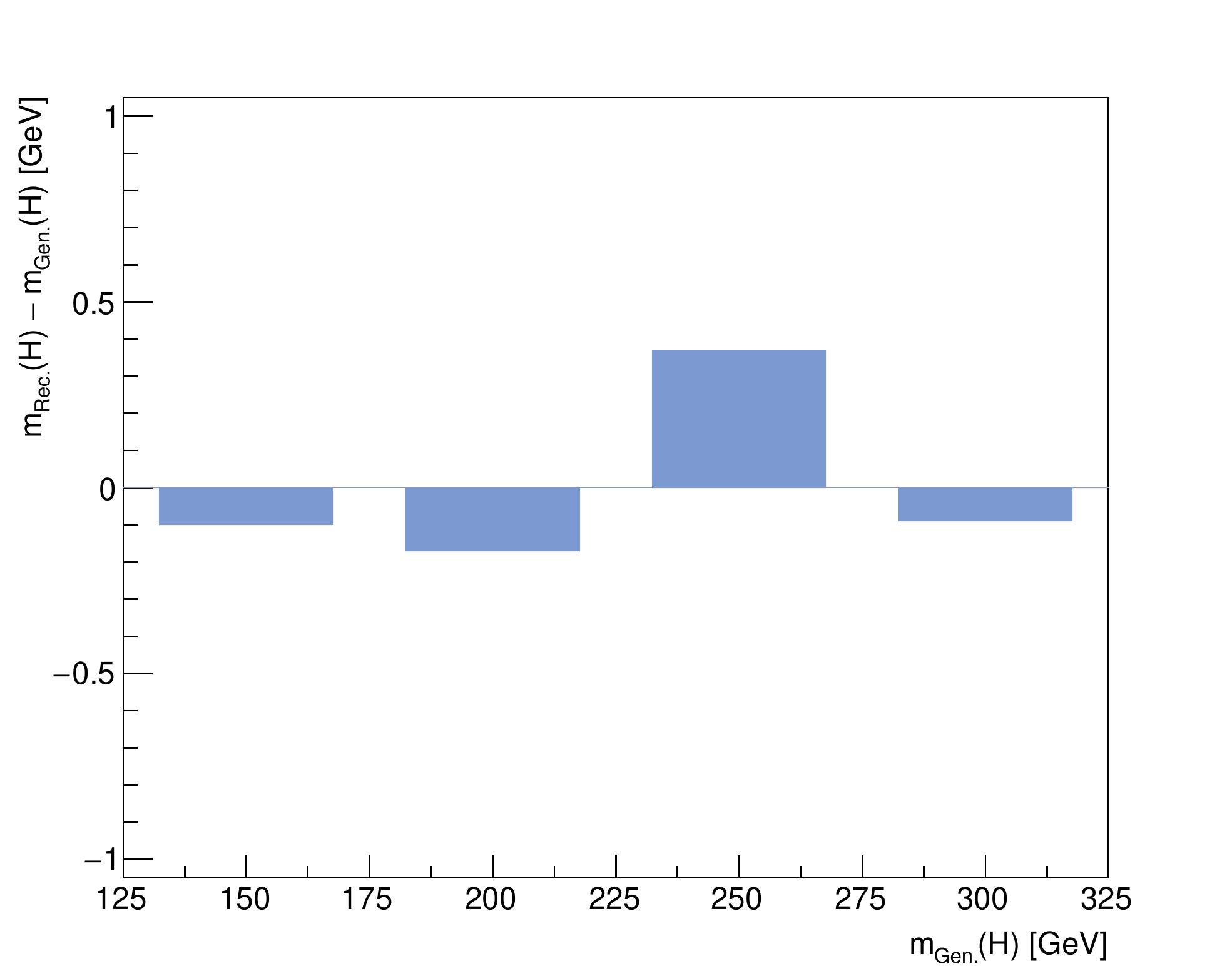}
  \caption{Reconstructed Higgs mass minus generated Higgs mass corresponding to the four assumed benchmark points after correction.}
\label{correctedMass}
\end{figure}

Aside from the mentioned error sources, a real experiment can be affected by other errors resulting from underlying-events, pile-up, electronic noise, etc. Hence an accurate correction to the distiguished jets and their properties is impossible unless all the corrections concerning underlying-events, pile-up, jet energy scale uncertainties, Data/MC calibration, etc., are taken into account. Since studying these effects lies beyond the scope of this analysis, a simple off-set correction is applied to make the reconstructed and generated Higgs masses matched as well as possible. This correction can be done by first using a flat function to fit to the plot of figure \ref{recmassMinusGenmass} and find the average difference between the reconstructed and generated masses, and then increasing the reconstructed masses by this average value. As seen in figure \ref{recmassMinusGenmass}, the average difference is $-0.33$ GeV which is used to perform the off-set correction. The obtained corrected masses are shown in table \ref{correctecrec.masstab} including fit uncertainties. Figure \ref{correctedMass} shows the difference between reconstructed and and generated Higgs masses after correction.

As seen in table \ref{correctecrec.masstab} and figure \ref{correctedMass}, the Higgs mass can be measured using the di-muon invariant mass distribution for all of the assumed benchmark points with few GeV uncertainty which is in fact a statistical error. However, in a real experiment there are some considerable sources of systematic errors such as particle momentum resolution, the jet energy scale and resolution, the $b$-tagging uncertainty and the uncertainty arising from the fit function used to obtain the probability density function (p.d.f.) of the distributions. The background modeling must be treated with special care for a reasonable observation of the signal. A thorough study and comparison of the distributions of different background samples resulting from real data and MC is needed to achieve a reasonable p.d.f. for the background. 
\begin{table}[h]
\normalsize
\fontsize{11}{7.2} 
    \begin{center}
         \begin{tabular}{ >{\centering\arraybackslash}m{1.9in} ? >{\centering\arraybackslash}m{.85in} ? >{\centering\arraybackslash}m{.85in} ?>{\centering\arraybackslash}m{.85in} ? >{\centering\arraybackslash}m{.85in} ? >{\centering\arraybackslash}m{.85in} ?} 
& \cellcolor{blizzardblue}{BP 1} & \cellcolor{blizzardblue}{BP 2} & \cellcolor{blizzardblue}{BP 3} & \cellcolor{blizzardblue}{BP 4} \parbox{0pt}{\rule{0pt}{1ex+\baselineskip}}\\ \Xhline{6\arrayrulewidth}
    \cellcolor{blizzardblue}{Generated mass [GeV]} & 150 & 200 & 250 & 300  \parbox{0pt}{\rule{0pt}{1ex+\baselineskip}}\\ \Xhline{2\arrayrulewidth}
    \cellcolor{blizzardblue}{Reconstructed mass [GeV]} & 149.9$\pm$1.77 & 199.83$\pm$1.51 & 250.37$\pm$1.54 & 299.91$\pm$1.81 \parbox{0pt}{\rule{0pt}{1ex+\baselineskip}}\\ \Xhline{2\arrayrulewidth}
        \end{tabular}
\caption{Generated and reconstructed Higgs masses with associated statistical errors. \label{correctecrec.masstab}}
  \end{center}
\end{table}
\section{Signal significance}
The signal observability is quantified through the signal significance calculation. Using the distributions of figures \ref{H150}-\ref{H300}, a mass window cut is determined for each benchmark point independently. Applying the mass window cuts, the final total efficiency, number of surviving signal and background events and their ratio and also the signal significance are obtained as shown in table \ref{resultstab}. The results shown in table \ref{resultstab} are obtained assuming $L=3000\ fb^{-1}$ and $\tan\beta=10$.
\begin{table}[h]
\normalsize
\fontsize{11}{7.2} 
    \begin{center}
         \begin{tabular}{ >{\centering\arraybackslash}m{1.9in} ? >{\centering\arraybackslash}m{.6in} ? >{\centering\arraybackslash}m{.6in} ?>{\centering\arraybackslash}m{.6in} ? >{\centering\arraybackslash}m{.6in} ? >{\centering\arraybackslash}m{.6in} ?} 
& \cellcolor{blizzardblue}{BP 1} & \cellcolor{blizzardblue}{BP 2} & \cellcolor{blizzardblue}{BP 3} & \cellcolor{blizzardblue}{BP 4} \parbox{0pt}{\rule{0pt}{1ex+\baselineskip}}\\ \Xhline{6\arrayrulewidth}
    \cellcolor{blizzardblue}{$m_{H}$} & 150 & 200 & 250 & 300  \parbox{0pt}{\rule{0pt}{1ex+\baselineskip}}\\ \Xhline{2\arrayrulewidth}
    \cellcolor{blizzardblue}{Mass window [GeV]} & 134-164 & 190-210 & 242-257 & 295-305  \parbox{0pt}{\rule{0pt}{1ex+\baselineskip}}\\ \Xhline{2\arrayrulewidth}
    \cellcolor{blizzardblue}{Total eff.} & 0.0031 & 0.0019 & 0.0014 & 0.0012  \parbox{0pt}{\rule{0pt}{1ex+\baselineskip}}\\ \Xhline{2\arrayrulewidth}
    \cellcolor{blizzardblue}{$S$} & 80 & 36 & 19 & 9   \parbox{0pt}{\rule{0pt}{1ex+\baselineskip}}\\ \Xhline{2\arrayrulewidth}
    \cellcolor{blizzardblue}{$B$} & 75 & 20 & 7 & 2  \parbox{0pt}{\rule{0pt}{1ex+\baselineskip}}\\ \Xhline{2\arrayrulewidth}
\cellcolor{blizzardblue}{$S/B$} & 1.1 & 1.8 & 2.7 & 4.3  \parbox{0pt}{\rule{0pt}{1ex+\baselineskip}}\\ \Xhline{2\arrayrulewidth}
\cellcolor{blizzardblue}{$S/\sqrt{B}$} & 9.2 & 8.0 & 7.2 & 6.4  \parbox{0pt}{\rule{0pt}{1ex+\baselineskip}}\\ \Xhline{2\arrayrulewidth}
  \end{tabular}
\caption{Higgs mass window cut, signal total efficiency, number of signal and background events after all selection cuts and mass window cut, signal to background ratio and signal significance. \label{resultstab}}
  \end{center}
\end{table}

According to table \ref{resultstab}, the signal significance decreases as the Higgs mass $m_H$ increases. This is not a surprising result though, as it is a consequence of the fact that the signal cross section decreases as the Higgs mass gets larger. As a result, the higher the Higgs mass, the harder the observation. 

Considering the results shown in table \ref{resultstab}, it is concluded that an observable signal (exceeding 5$\sigma$) can be extracted from the di-muon invariant mass distribution for any one of the four benchmark points at 3000 $fb^{-1}$.

\section{Conclusions}
The observability of a neutral CP-even Higgs boson $H$ at a linear collider operating at $\sqrt{s}=1$ TeV was studied in the framework of a type \RN{4} 2HDM. The signal process was assumed to be $e^- e^+ \rightarrow A H \rightarrow ZHH$ followed by hadronic (leptonic) decay of $Z$ ($H$) bosons. Four benchmark points were hypothesized and the simulation was performed for each one independently. The Higgs mass $m_H$ range under study was 150 GeV to 300 GeV in increments of 50 GeV. Although the branching ratio of the Higgs boson $H$ decay to a pair of muons is very small, the ability to accurately identify the muons compensates for the branching ratio smallness and plays a significant role in this study. Taking advantage of the kinematic differences between signal and background events, appropriate selection cuts were applied and the Higgs boson candidate mass distribution was obtained. As the Higgs mass $m_H$ gets heavier the signal cross section decreases and this fact could have caused a major obstacle to observing the Higgs boson. However, the weakness of the background tail partially compensated for the decrease in the signal cross section and helped observation of the Higgs boson $H$ with masses up to 300 GeV. These results indicate that for all of the four assumed benchmark points, an observable signal can be extracted from the SM background with mass measurement possibility at 3000 $fb^{-1}$.

\section*{Acknowledgements}
The analysis presented in this work was fully performed using the computing cluster at Shiraz University, college of sciences. We would like to thank Dr. Mogharrab for his careful maintenance and operation of the computing cluster. 


\begin{thebibliography}{10}
\bibliographystyle{JHEP}  

\bibitem{HiggsObservationCMS}
{\bf CMS} Collaboration, S.~Chatrchyan et~al., {\it {Observation of a new boson
  at a mass of 125 GeV with the CMS experiment at the LHC}},  {\em Phys. Lett.}
  {\bf B716} (2012) 30--61, [\href{http://arxiv.org/abs/1207.7235}{{\tt
  arXiv:1207.7235}}].

\bibitem{HiggsObservationATLAS}
{\bf ATLAS} Collaboration, G.~Aad et~al., {\it {Observation of a new particle
  in the search for the Standard Model Higgs boson with the ATLAS detector at
  the LHC}},  {\em Phys. Lett.} {\bf B716} (2012) 1--29,
  [\href{http://arxiv.org/abs/1207.7214}{{\tt arXiv:1207.7214}}].

\bibitem{SM1}
R.~Brout and F.~Englert, {\it {Spontaneous symmetry breaking in gauge theories:
  A Historical survey}},  in {\em {High-energy physics. Proceedings,
  International Europhysics Conference, Jerusalem, Israel, August 19-25,
  1997}}, pp.~3--10, 1998.
\newblock \href{http://arxiv.org/abs/hep-th/9802142}{{\tt hep-th/9802142}}.

\bibitem{SM2}
P.~W. Higgs, {\it {Broken Symmetries and the Masses of Gauge Bosons}},  {\em
  Phys. Rev. Lett.} {\bf 13} (1964) 508--509.

\bibitem{SM3}
P.~W. Higgs, {\it {Broken symmetries, massless particles and gauge fields}},
  {\em Phys. Lett.} {\bf 12} (1964) 132--133.

\bibitem{SM4}
G.~S. Guralnik, C.~R. Hagen, and T.~W.~B. Kibble, {\it {Global Conservation
  Laws and Massless Particles}},  {\em Phys. Rev. Lett.} {\bf 13} (1964)
  585--587.

\bibitem{SM5}
P.~W. Higgs, {\it {Spontaneous Symmetry Breakdown without Massless Bosons}},
  {\em Phys. Rev.} {\bf 145} (1966) 1156--1163.

\bibitem{SM6}
T.~W.~B. Kibble, {\it Symmetry breaking in non-abelian gauge theories},  {\em
  Phys. Rev.} {\bf 155} (Mar, 1967) 1554--1561.

\bibitem{MSSM1}
I.~J.~R. Aitchison, {\it {Supersymmetry and the MSSM: An Elementary
  introduction}},  \href{http://arxiv.org/abs/hep-ph/0505105}{{\tt
  hep-ph/0505105}}.

\bibitem{axionModels}
J.~E. Kim, {\it Light pseudoscalars, particle physics and cosmology},  {\em
  Physics Reports} {\bf 150} (1987), no.~1 1 -- 177.

\bibitem{SMinabilityToExplainBaryonAsymmetry}
M.~Trodden, {\it {Electroweak baryogenesis: A Brief review}},  in {\em
  {Proceedings, 33rd Rencontres de Moriond 98 electrowek interactions and
  unified theories: Les racs, France, Mar 14-21, 1998}}, pp.~471--480, 1998.
\newblock \href{http://arxiv.org/abs/hep-ph/9805252}{{\tt hep-ph/9805252}}.

\bibitem{2hdm_TheoryPheno}
G.~C. Branco, P.~M. Ferreira, L.~Lavoura, M.~N. Rebelo, M.~Sher, and J.~P.
  Silva, {\it {Theory and phenomenology of two-Higgs-doublet models}},  {\em
  Phys. Rept.} {\bf 516} (2012) 1--102,
  [\href{http://arxiv.org/abs/1106.0034}{{\tt arXiv:1106.0034}}].

\bibitem{2hdm1}
T.~D. Lee, {\it {A Theory of Spontaneous T Violation}},  {\em Phys. Rev.} {\bf
  D8} (1973) 1226--1239.

\bibitem{2hdm2}
S.~L. Glashow and S.~Weinberg, {\it {Natural Conservation Laws for Neutral
  Currents}},  {\em Phys. Rev.} {\bf D15} (1977) 1958.

\bibitem{2hdm3}
G.~C. Branco, {\it {Spontaneous {CP} Nonconservation and Natural Flavor
  Conservation: A Minimal Model}},  {\em Phys. Rev.} {\bf D22} (1980) 2901.

\bibitem{2hdm4_CompositeHiggs}
J.~Mrazek, A.~Pomarol, R.~Rattazzi, M.~Redi, J.~Serra, and A.~Wulzer, {\it {The
  Other Natural Two Higgs Doublet Model}},  {\em Nucl. Phys.} {\bf B853} (2011)
  1--48, [\href{http://arxiv.org/abs/1105.5403}{{\tt arXiv:1105.5403}}].

\bibitem{2hdm_HiggsSector1}
S.~Davidson and H.~E. Haber, {\it {Basis-independent methods for the
  two-Higgs-doublet model}},  {\em Phys. Rev.} {\bf D72} (2005) 035004,
  [\href{http://arxiv.org/abs/hep-ph/0504050}{{\tt hep-ph/0504050}}]. [Erratum:
  Phys. Rev.D72,099902(2005)].

\bibitem{2hdm_HiggsSector2}
M.~Aoki, S.~Kanemura, K.~Tsumura, and K.~Yagyu, {\it {Models of Yukawa
  interaction in the two Higgs doublet model, and their collider
  phenomenology}},  {\em Phys. Rev.} {\bf D80} (2009) 015017,
  [\href{http://arxiv.org/abs/0902.4665}{{\tt arXiv:0902.4665}}].

\bibitem{tanbsignificance}
H.~E. Haber and D.~O'Neil, {\it {Basis-independent methods for the
  two-Higgs-doublet model. II. The Significance of tan$\beta$}},  {\em Phys.
  Rev.} {\bf D74} (2006) 015018,
  [\href{http://arxiv.org/abs/hep-ph/0602242}{{\tt hep-ph/0602242}}]. [Erratum:
  Phys. Rev.D74,no.5,059905(2006)].

\bibitem{MSSM2}
E.~Ma and D.~Ng, {\it {New supersymmetric option for two Higgs doublets}},
  {\em Phys. Rev.} {\bf D49} (1994) 6164--6167,
  [\href{http://arxiv.org/abs/hep-ph/9305230}{{\tt hep-ph/9305230}}].

\bibitem{MSSM3}
A.~Djouadi, {\it {The Anatomy of electro-weak symmetry breaking. II. The Higgs
  bosons in the minimal supersymmetric model}},  {\em Phys. Rept.} {\bf 459}
  (2008) 1--241, [\href{http://arxiv.org/abs/hep-ph/0503173}{{\tt
  hep-ph/0503173}}].

\bibitem{MSSM4}
C.~Csaki, {\it {The Minimal supersymmetric standard model (MSSM)}},  {\em Mod.
  Phys. Lett.} {\bf A11} (1996) 599,
  [\href{http://arxiv.org/abs/hep-ph/9606414}{{\tt hep-ph/9606414}}].

\bibitem{Kanemura}
S.~Kanemura, K.~Tsumura, and H.~Yokoya, {\it {Multi-tau-lepton signatures at
  the LHC in the two Higgs doublet model}},  {\em Phys. Rev.} {\bf D85} (2012)
  095001, [\href{http://arxiv.org/abs/1111.6089}{{\tt arXiv:1111.6089}}].

\bibitem{naturalflavorcons.}
E.~A. Paschos, {\it {Diagonal Neutral Currents}},  {\em Phys. Rev.} {\bf D15}
  (1977) 1966.

\bibitem{Barger_2hdmTypes}
V.~D. Barger, J.~L. Hewett, and R.~J.~N. Phillips, {\it {New Constraints on the
  Charged Higgs Sector in Two Higgs Doublet Models}},  {\em Phys. Rev.} {\bf
  D41} (1990) 3421--3441.

\bibitem{2HDMTypeI_LHC}
G.~C. Dorsch, S.~J. Huber, K.~Mimasu, and J.~M. No, {\it {Echoes of the
  Electroweak Phase Transition: Discovering a second Higgs doublet through $A_0
  \rightarrow ZH_0$}},  {\em Phys. Rev. Lett.} {\bf 113} (2014), no.~21 211802,
  [\href{http://arxiv.org/abs/1405.5537}{{\tt arXiv:1405.5537}}].

\bibitem{2hdmtype4-1}
R.~M. Barnett, G.~Senjanovic, L.~Wolfenstein, and D.~Wyler, {\it {Implications
  of a Light Higgs Scalar}},  {\em Phys. Lett.} {\bf B136} (1984) 191--195.

\bibitem{2hdmtype4-2}
R.~M. Barnett, G.~Senjanovic, and D.~Wyler, {\it {Tracking Down Higgs Scalars
  With Enhanced Couplings}},  {\em Phys. Rev.} {\bf D30} (1984) 1529.

\bibitem{L2HDMLHC}
S.~Su and B.~Thomas, {\it {The LHC Discovery Potential of a Leptophilic
  Higgs}},  {\em Phys. Rev.} {\bf D79} (2009) 095014,
  [\href{http://arxiv.org/abs/0903.0667}{{\tt arXiv:0903.0667}}].

\bibitem{2HDMrhoConstrainingMeasurement}
{\bf Particle Data Group} Collaboration, W.~M. Yao et~al., {\it {Review of
  Particle Physics}},  {\em J. Phys.} {\bf G33} (2006) 1--1232.

\bibitem{2HDMrhoConstraint-1}
S.~Bertolini, {\it Quantum effects in a two higgs doublet model of the
  electroweak interactions},  {\em Nuclear Physics B} {\bf 272} (1986), no.~1
  77 -- 98.

\bibitem{2HDMrhoConstraint-2}
A.~Denner, R.~Guth, and J.~Kühn, {\it Relaxation of top mass limits in the
  two-higgs-doublet model},  {\em Physics Letters B} {\bf 240} (1990),
  no.~3–4 438 -- 440.

\bibitem{2HDMdeltaRho}
W.~Grimus, L.~Lavoura, O.~M. Ogreid, and P.~Osland, {\it {A Precision
  constraint on multi-Higgs-doublet models}},  {\em J. Phys.} {\bf G35} (2008)
  075001, [\href{http://arxiv.org/abs/0711.4022}{{\tt arXiv:0711.4022}}].

\bibitem{2HDMHiggsPotentialPositivity}
N.~G. Deshpande and E.~Ma, {\it Pattern of symmetry breaking with two higgs
  doublets},  {\em Phys. Rev. D} {\bf 18} (Oct, 1978) 2574--2576.

\bibitem{2HDMtreeLevelUnitarity-1}
H.~Huffel and G.~Pocsik, {\it {Unitarity Bounds on Higgs Boson Masses in the
  {Weinberg-Salam} Model With Two Higgs Doublets}},  {\em Z. Phys.} {\bf C8}
  (1981) 13.

\bibitem{2HDMtreeLevelUnitarity-2}
J.~Maalampi, J.~Sirkka, and I.~Vilja, {\it Tree level unitarity and triviality
  bounds for two-higgs models},  {\em Physics Letters B} {\bf 265} (1991),
  no.~3 371 -- 376.

\bibitem{2HDMtreeLevelUnitarity-3}
S.~Kanemura, T.~Kubota, and E.~Takasugi, {\it {Lee-Quigg-Thacker bounds for
  Higgs boson masses in a two doublet model}},  {\em Phys. Lett.} {\bf B313}
  (1993) 155--160, [\href{http://arxiv.org/abs/hep-ph/9303263}{{\tt
  hep-ph/9303263}}].

\bibitem{2HDMtreeLevelUnitarity-4}
A.~G. Akeroyd, A.~Arhrib, and E.-M. Naimi, {\it {Note on tree level unitarity
  in the general two Higgs doublet model}},  {\em Phys. Lett.} {\bf B490}
  (2000) 119--124, [\href{http://arxiv.org/abs/hep-ph/0006035}{{\tt
  hep-ph/0006035}}].

\bibitem{2hdmc1}
D.~Eriksson, J.~Rathsman, and O.~Stal, {\it {2HDMC: Two-Higgs-Doublet Model
  Calculator Physics and Manual}},  {\em Comput. Phys. Commun.} {\bf 181}
  (2010) 189--205, [\href{http://arxiv.org/abs/0902.0851}{{\tt
  arXiv:0902.0851}}].

\bibitem{2hdmc2}
D.~Eriksson, J.~Rathsman, and O.~Stal, {\it {2HDMC: Two-Higgs-doublet model
  calculator}},  {\em Comput. Phys. Commun.} {\bf 181} (2010) 833--834.

\bibitem{lep1}
{\bf ALEPH} Collaboration, R.~Barate et~al., {\it {Search for charged Higgs
  bosons in e+ e- collisions at energies up to $\sqrt{s} = 189$ GeV}},  {\em
  Phys. Lett.} {\bf B487} (2000) 253--263,
  [\href{http://arxiv.org/abs/hep-ex/0008005}{{\tt hep-ex/0008005}}].

\bibitem{lep2}
{\bf L3} Collaboration, M.~Acciarri et~al., {\it {Search for charged Higgs
  bosons in $e^{+} e^{-}$ collisions at center center-of-mass energies up to
  202-GeV}},  {\em Phys. Lett.} {\bf B496} (2000) 34--42,
  [\href{http://arxiv.org/abs/hep-ex/0009010}{{\tt hep-ex/0009010}}].

\bibitem{lepexclusion1}
{\bf OPAL, DELPHI, L3, ALEPH, LEP Higgs Working Group for Higgs boson searches}
  Collaboration, {\it {Search for charged Higgs bosons: Preliminary combined
  results using LEP data collected at energies up to 209-GeV}},  in {\em
  {Lepton and photon interactions at high energies. Proceedings, 20th
  International Symposium, LP 2001, Rome, Italy, July 23-28, 2001}}.

\bibitem{lepexclusion2}
{\bf DELPHI, OPAL, ALEPH, LEP Higgs Working Group, L3} Collaboration, {\it
  {Searches for the neutral Higgs bosons of the MSSM: Preliminary combined
  results using LEP data collected at energies up to 209-GeV}},  in {\em
  {Lepton and photon interactions at high energies. Proceedings, 20th
  International Symposium, LP 2001, Rome, Italy, July 23-28, 2001}}.

\bibitem{CMSNeutralHiggs}
{\bf CMS} Collaboration, {\it {Search for a neutral MSSM Higgs boson decaying
  into $\tau\tau$ with $12.9~\mathrm{fb}^{-1}$ of data at
  $\sqrt{s}=13~\mathrm{TeV}$}},  {\em CMS Collaboration, CMS-PAS-HIG-16-037}.

\bibitem{ATLASneutralHiggs}
{\bf ATLAS} Collaboration {\em ATLAS-CONF-2016-085}.

\bibitem{Misiak}
M.~Misiak et~al., {\it {Updated NNLO QCD predictions for the weak radiative
  B-meson decays}},  {\em Phys. Rev. Lett.} {\bf 114} (2015), no.~22 221801,
  [\href{http://arxiv.org/abs/1503.01789}{{\tt arXiv:1503.01789}}].

\bibitem{pythia8.2}
T.~Sjöstrand, S.~Ask, J.~R. Christiansen, R.~Corke, N.~Desai, P.~Ilten,
  S.~Mrenna, S.~Prestel, C.~O. Rasmussen, and P.~Z. Skands, {\it {An
  Introduction to PYTHIA 8.2}},  {\em Comput. Phys. Commun.} {\bf 191} (2015)
  159--177, [\href{http://arxiv.org/abs/1410.3012}{{\tt arXiv:1410.3012}}].

\bibitem{fastjet1}
M.~Cacciari, {\it {FastJet: A Code for fast $k_t$ clustering, and more}},  in
  {\em {Deep inelastic scattering. Proceedings, 14th International Workshop,
  DIS 2006, Tsukuba, Japan, April 20-24, 2006}}, pp.~487--490, 2006.
\newblock \href{http://arxiv.org/abs/hep-ph/0607071}{{\tt hep-ph/0607071}}.
\newblock [,125(2006)].

\bibitem{fastjet2}
M.~Cacciari, G.~P. Salam, and G.~Soyez, {\it {FastJet User Manual}},  {\em Eur.
  Phys. J.} {\bf C72} (2012) 1896, [\href{http://arxiv.org/abs/1111.6097}{{\tt
  arXiv:1111.6097}}].

\bibitem{antikt}
M.~Cacciari, G.~P. Salam, and G.~Soyez, {\it {The Anti-k(t) jet clustering
  algorithm}},  {\em JHEP} {\bf 04} (2008) 063,
  [\href{http://arxiv.org/abs/0802.1189}{{\tt arXiv:0802.1189}}].

\bibitem{cliccdr}
L.~Linssen, A.~Miyamoto, M.~Stanitzki, and H.~Weerts, {\it {Physics and
  Detectors at CLIC: CLIC Conceptual Design Report}},
  \href{http://arxiv.org/abs/1202.5940}{{\tt arXiv:1202.5940}}.

\bibitem{root}
R.~Brun and F.~Rademakers, {\it {ROOT: An object oriented data analysis
  framework}},  {\em Nucl. Instrum. Meth.} {\bf A389} (1997) 81--86.

\end{thebibliography}

\providecommand{\href}[2]{#2}\begingroup\raggedright\endgroup


\end{document}